\documentclass[10pt,journal,cspaper,compsoc]{IEEEtran}
\usepackage{graphicx}
\usepackage{subfig}
\usepackage[algoruled,linesnumbered]{algorithm2e}
\usepackage{amsmath,amsthm, amssymb}
\usepackage{color}
\usepackage{hyperref}
\newtheorem{definition}{Definition}
\newtheorem{theorem}{Theorem}
\newenvironment{myproof}[1][\proofname]{\par\noindent{\bfseries #1. }}{\qed}{\endproof}
\newcommand{\be}{\begin{eqnarray}}
\newcommand{\ee}{\end{eqnarray}}
\newcommand{\ben}{\begin{eqnarray*}}
\newcommand{\een}{\end{eqnarray*}}

\newcommand{\mcn}{\mathcal{N}}
\newcommand{\lam}{\lambda}
\newcommand{\sig}{\sigma}
\newcommand{\e}{\mathrm{\mathbf{E}}}

\begin{document}
\title{Priority-Aware Private Matching Schemes for Proximity-Based Mobile Social Networks}
\author{Ben~Niu,
        Tanran~Zhang,
        Xiaoyan~Zhu,
        Hui~Li
        and~Zongqing~Lu
\IEEEcompsocitemizethanks{
\IEEEcompsocthanksitem  B. Niu, X. Zhu and H. Li are with the School of Telecommunications Engineering, Xidian University, 710071, China. E-mail: xd.niuben@gmail.com and \{xyzhu, lihui\}@mail.xidian.edu.cn.
\IEEEcompsocthanksitem T. Zhang is with GSIS, Tohoku University, Sendai, Japan. E-mail: xubu3@163.com.
\IEEEcompsocthanksitem  Z. Lu is with the Department of Computer Science and Engineering, The Pennsylvania State University, University Park, PA 16802. E-mail: zongqing@cse.psu.edu.}
}

\markboth{}%
{Shell \MakeLowercase{\textit{et al.}}: Bare Demo of IEEEtran.cls for Computer Society Journals}
\IEEEcompsoctitleabstractindextext{%
\begin{abstract}
The rapid developments of mobile devices and online social networks have resulted in increasing attention to Mobile Social Networking (MSN). The explosive growth of mobile-connected and location-aware devices makes it possible and meaningful to do the Proximity-based Mobile Social Networks (PMSNs). Users can discover and make new social interactions easily with physical-proximate mobile users through WiFi/Bluetooth interfaces embedded in their smartphones. However, users enjoy these conveniences at the cost of their growing privacy concerns. To address this problem, we propose a suit of priority-aware private matching schemes to privately match the similarity with potential friends in the vicinity. Unlike most existing work, our proposed priority-aware matching scheme (\emph{P-match}) achieves the privacy goal by combining the commutative encryption function and the Tanimoto similarity coefficient which considers both the number of common attributes between users as well as the corresponding priorities on each common attribute. Further, based on the newly constructed similarity function which takes the ratio of attributes matched over all the input set into consideration, we design an enhanced version to deal with some potential attacks such as unlimitedly inputting the attribute set on either the \emph{initiator} side or the \emph{responder} side, etc. Finally, our proposed \emph{E-match} avoids the heavy cryptographic operations and improves the system performance significantly by employing a novel use of the Bloom filter. The security and communication/computation overhead of our schemes are thoroughly analyzed and evaluated via detailed simulations and implementation.
\end{abstract}
}
\maketitle
\IEEEdisplaynotcompsoctitleabstractindextext
\section{Introduction}
\label{sec_intro}
\IEEEPARstart{S}{ocial} networking is one of the fastest-growing activities among mobile users domestically and worldwide. According to eMarketer \cite{Emarketer11}, they estimate the number of US smartphone users will reach 192.4 millions by 2016, and 2.28 billions worldwide. By smartphones equipped with WiFi/Bluetooth modules, users can communicate with others easily and exchange information, content and media on shared communities such as Facebook or Foursquare. Among these services, an important classification is Proximity-based Mobile Social Networks (PMSNs), which deeply relies on mobile users' physical proximity. This kind of applications can provide us more opportunities to discover and make new social interactions within some public places, such as airports, bars or other social spots. PMSNs thus gain increasing attention in social networking.

Normally, to enjoy these activities, people always need to reveal some information such as their attributes or personal information to potential friends nearby as the first step. A straightforward way is that, an \emph{initiator} broadcasts her attributes to nearby users directly, and the \emph{responder}s decide whether to contact her based on common attributes. Obviously, the user's privacy is revealed during such process. Since some of these attributes may be sensitive or private to the user, it is harmful to leak them to everyone nearby, especially the potential malicious users.

To address this problem, many research solutions \cite{serendipity05, SmokeScreen07, Smile09, MobiClique09, Secure11, Lu11, FindU11, FineG12, Ben13G} have been proposed. Among these solutions, most of them employ third party servers, and thus, they become the bottlenecks from both security$/$privacy and system performance points of view. Although this kind of third party servers can be set to offline, the mobile users need to access them to register identities and obtain the matching results, which bring extra 3G/4G communication cost. Some researchers consider this situation as a Private Set Intersection (PSI) problem \cite{PSI04, Kissner05, Sang09, Ateniese11} and try to achieve private matching while avoiding the third party servers by employing Secure Multi-party Computation (SMC) \cite{FindU11} and Paillier Cryptosystem \cite{Secure11, FineG12}. Unfortunately, the PSI-based solutions can avoid the trusted server effectively but always fail to improve the system performance due to the heavy cryptographic operations.

\begin{figure}[!t]
  \centering
  \subfloat[Scenario 1]{\includegraphics[width=2.67in]{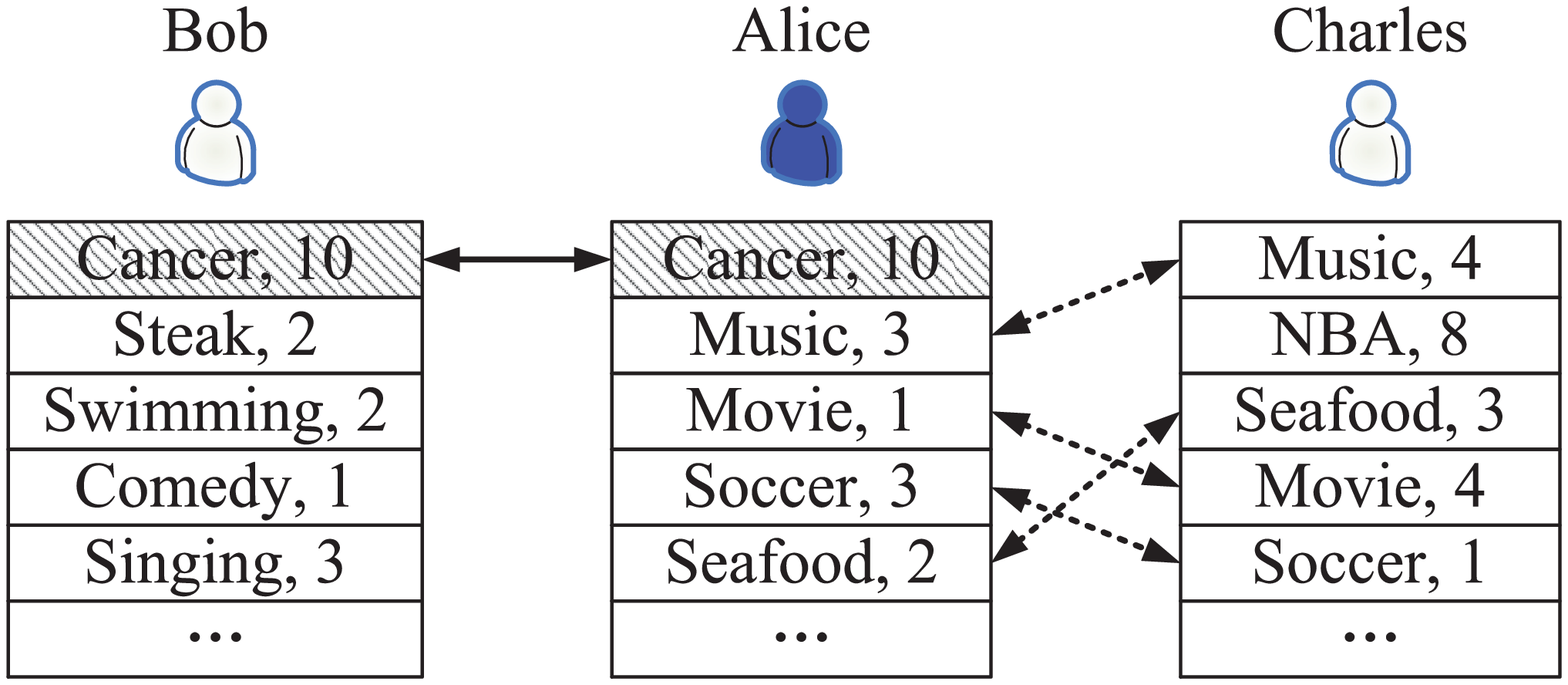}
    \label{subfig_motivation1}}
    \\
  \subfloat[Scenario 2]{\includegraphics[width=2.67in]{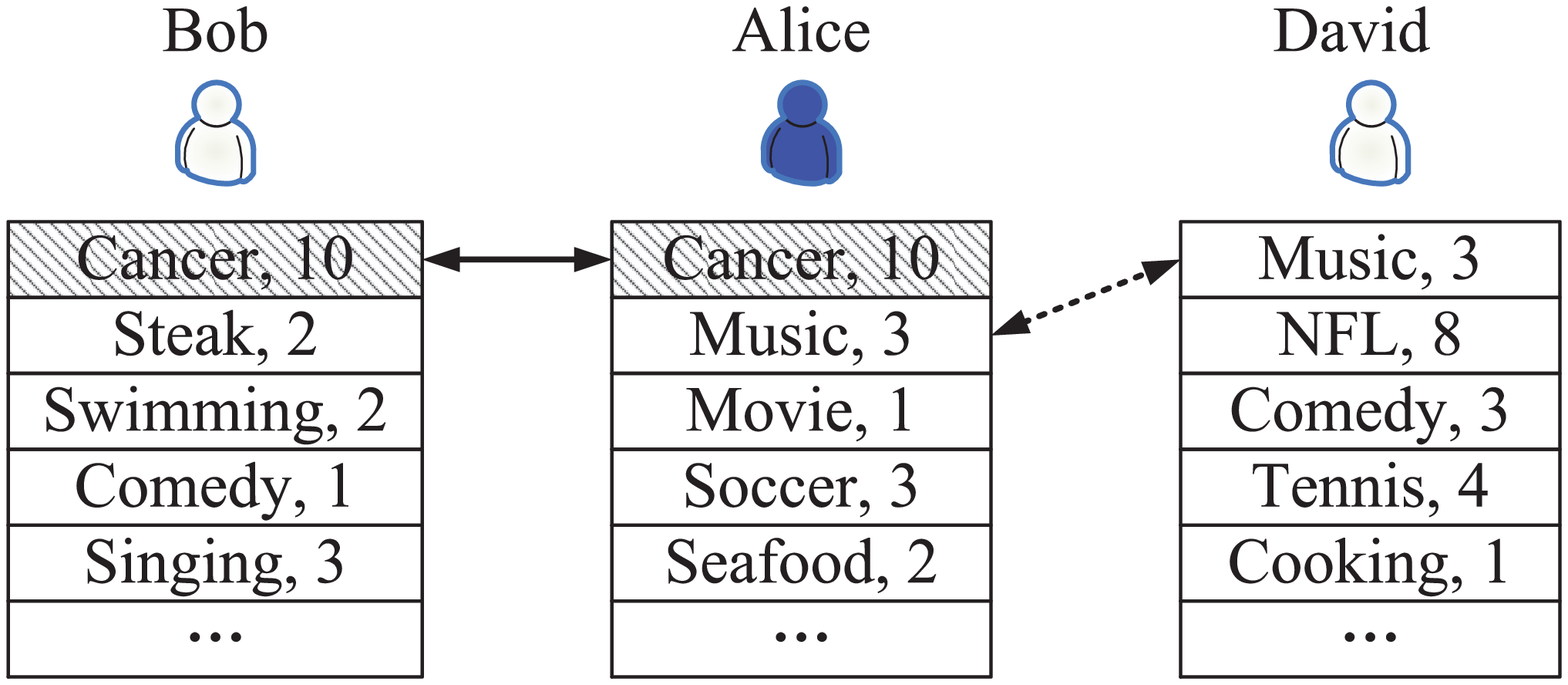}
    \label{subfig_motivation2}}
  \caption{Motivation}
  \label{fig_motivation}
\end{figure}

Moreover, existing schemes do not always produce accurate matching results. For example, \cite{FindU11} and \cite{Esmalltalker10} measure the similarity between users by simply counting the number of common attributes, and the matching decisions is made by checking whether the proximity measurement of two profiles is larger, equal, or smaller than a pre-defined threshold value in \cite{Xiaohui13}. However, in reality, user interests may be associated with different priorities. We illustrate our concerns with an example in Fig. \ref{fig_motivation}, the two-tuple represents user's interests with the corresponding priorities. The scenario is that \emph{Alice}'s father is suffering from cancer and she needs to go to hospital everyday after work. In this situation, the most interested person she wants to know is someone who is facing the same situation. Therefore, in the scenario shown in Fig. \ref{subfig_motivation1}, the potential friend of \emph{Alice} is \emph{Bob} instead of \emph{Charles}, even though there are four common attributes between \emph{Alice} and \emph{Charles} and only one with \emph{Bob}. However, this kind of schemes suffers the attack of choosing the attributes as many as possible on the adversary side, leading to the exposure of users' personal information. Although, the authors in \cite{FindU11} proposed a solution by limiting the number of input attributes (e.g., 200) to avoid this attack, it is hard to define a proper number of input attributes for different users. To achieve a fine-grained private matching, Zhang \emph{et al.} \cite{FineG12} consider the priority on each common attribute and define several privacy levels in their work, but they only pay attention on the difference of priorities on each common attribute between users, and ignore the priority value itself. That is to say, it works well for most cases except one shown in the scenario of Fig. \ref{subfig_motivation2}. \emph{Alice} has one common attribute with either \emph{Bob} or \emph{David}, it makes sense that she prefers to make friends with \emph{Bob} since she pays more attention on the attribute "cancer". However, the approach in \cite{FineG12} cannot differentiate \emph{David} and \emph{Bob} from the \emph{Alice}'s point of view, since \emph{Alice} and \emph{Bob} have the common attribute of "cancer" at priority 10, and \emph{Alice} and \emph{David} have the common attribute of "music" at priority 3.

From the aforementioned analysis, it is clear that existing work either rely on third party servers, or employ heavy cryptographic tools, or do not fully consider the users' privacy in terms of the priority assigned on each attribute, or produce inaccurate matching results. In this paper, we propose a set of priority-aware private matching schemes to accelerate the widely used PMSNs. The main contributions of this paper are shown as follows.

$\bullet$ We propose a set of schemes to achieve private matching for different privacy goals. \emph{P-match} achieves the priority-aware private matching with considering both the number of common attributes and the corresponding priorities. In the enhanced version \emph{P-match$^+$}, we construct a priority-aware Ochiai similarity coefficient to consider the ratio of attributes matched over all the input set in our similarity function, which can effectively prevent several attacks, such as unlimitedly inputting users' attribute sets. Finally, to make the private matching process more efficiently, we propose \emph{E-match}, which improves the system performance by avoiding the heavy commutative encryption function.

$\bullet$ We provide theoretic and experimental evidences that our proposed matching schemes are secure and can achieve our privacy goals. In addition, they are quite efficient compared to many existing work in terms of the computation cost, communication cost and energy consumption.

$\bullet$ We implement the proposed schemes into smartphones. The experimental results indicate the effectiveness and efficiency of our work.

The rest of this paper is organized as follows. Section \ref{sec_relat} reviews the related work. Section \ref{sec_preli} presents the preliminaries. Section \ref{sec_schemes} describes the designs of our proposed schemes. Section \ref{sec_secur} and \ref{sec_evalu} show the security analysis and the evaluation results. Finally, we conclude the paper in Section \ref{sec_concl}.

\section{Related Work}
\label{sec_relat}
There is a series of applications to provide private matching between users in PMSNs. Most of these solutions employ third party servers, which are always trusted and acting as matching centers to serve users. Specifically, each user sends her attributes to the server, the server replies users with the matching result to indicate the potential "friends". The servers need to know the users' personal information to perform the matching process, it is thus much dangerous when the servers are compromised. Social serendipity \cite{serendipity05} provided mobile users more opportunities to make social interactions with potential friends nearby. However, it deeply relied on a trusted server, which keeps all users' profiles and computes the similarity between users when needed. The authors in \cite{SmokeScreen07} improved this problem by replacing the trusted server with a service provider, such as Facebook. However, users' profiles are exchanged in plaintext, which lead to serious privacy leakage. SmokeScreen \cite{Manweiler09} introduced opaque identifier into the information exchanging phase to protect user's real identity. It employed a broker, which also acts as a trusted server to provide matching results to users. As a result, the broker known who is interested in solving whose opaque identifier, which means the broker can infer the relationship between users with high possibility. In their follow-on work \cite{Smile09}, they avoided this problem by not disclosing the personal information for matching, and using the location and time information as a replacement. Unfortunately, this kind of servers are still bottlenecks. The servers still need to know these information to perform the matching process. To avoid the third party servers, many cryptographic tools-based solutions have been proposed over recent years. Some researchers conclude this situation into the Private Set Intersection (PSI) or Authorized PSI (APSI) problem \cite{PSI04, Kissner05, Sang09, Ateniese11}, which can effectively avoid the third party servers. Based on PSI, Li \emph{et al.} \cite{FindU11} proposed a set of privacy-preserving profile matching schemes, where an initiating user can find the best match with minimal information leakage to others based on the security properties of Secure Multi-party Computation (SMC). However, the expensive computation cost brought by the heavy cryptographic operations decrease the system performance significantly. In addition, their schemes may fail in reality due to the ignorance on the priorities assigned on each attributes. Zhang \emph{et al.} presented a set of fine-grained private matching schemes to achieve the requirements of mobile users in reality. Under the protection of Paillier Cryptosystem, their schemes achieved fine-grained private matching with considering both the number of common attributes and the assigned priorities. However, they only paid attention on the differences of the priorities.

Agrawal \emph{et al.} \cite{Agrawal03} proposed a privacy-preserving protocol by using the commutative encryption function, which is more lightweight, to realize secret information sharing between users. The keyed hash functions are also employed to protect the sensitive attributes for mobile users. Then, Vaidya \emph{et al.} \cite{Vaidya05} extended the secret information sharing phase into N-party setting, and Veneta \emph{et al.} \cite{veneta08} implemented this idea to detect friend-of-friend in mobile social networks. However, since the inherent weaknesses of the scheme in \cite{Agrawal03}, such as unlimitedly inputting behavior and lying behavior, simply employing the commutative encryption function-based solution cannot provide thorough security and privacy properties. In this paper, we provide more properties by combining commutative encryption function with other techniques, such as similarity functions.

\section{Preliminaries}
\label{sec_preli}
In this section, we first state the problem and give the adversary models. Then, we describe the design goals and the cryptography tools in this paper.

\subsection{Problem Statement}
\label{subsec_statement}
In PMSNs, each user holds a profile with two dimensional vectors, $U = \{\langle u_1, v_1 \rangle, \langle u_2, v_2 \rangle, \cdots, \langle u_m, v_m \rangle\}$, where $u_i$ represents the user's attribute and $v_i$ means the priority value assigned to this particular attribute by the user, such as $v_i = 1, \ldots, 9$. Normally, bigger value indicates higher priority.
Given a user and her profile $U$, the problem is how to find the best matched friend for the user with privacy-preserving based on the criteria that the best matched friend should have more common attributes with the user, especially the same priorities on the attributes.

\subsection{Adversary Models}
\label{subsec_adversary}
Since cryptography tools such as Public Key Infrastructure (PKI) can be easily implemented to protect current communication systems, the attacks from outside adversaries, such as eavesdropping the wireless communication channels or modifying, replying and injecting the captured messages can be easily prevented. Therefore, in this paper, we assume the \emph{initiator} is honest-but-curious \cite{PSI04} or even an attacker directly. That means the \emph{initiator} will honestly follow the protocols, but tries to learn more information than allowed in the honest-but-curious model; or the \emph{initiator} can directly be an attacker, who may illegally input his attributes and the related priorities, to learn more information of nearby users. We also assume that the \emph{responder} is legal or honest-but-curious with two reasons: the identity of a \emph{responder} can be easily authorized by another authority entity, i.e., office in the hospital in the example mentioned in Sec. \ref{sec_intro}; another reason is that we can reverse our protocols to achieve a dual matching. We further assume that neither the \emph{initiator} nor the \emph{responder} can modify and obtain the result of parameters in the running protocols.

\subsection{Design Goals}
\label{subsec_goals}
Our main goal is to thwart the aforementioned threats from either \emph{initiator} or \emph{responder} side. According to the amount of information disclosed during the protocols execution, we define two privacy levels from the \emph{initiator} \emph{Alice}'s point of view, which can also be equivalently defined from \emph{Bob}'s viewpoint.

\begin{definition} [Privacy Level I]
\label{def_level1}
When the protocol ends, \emph{Bob} learns the set of common attributes with \emph{Alice}, as well as priorities on these common attributes, and \emph{Alice} learns the similarity value.
\end{definition}
In this case, the \emph{initiator} believes that all the nearby users are legal or honest-but-curious, and they can be authorized by another entity. For instance, \emph{Alice} may be a new patient in the hospital ward. Her aim is to find a best match.
\begin{definition} [Privacy Level II]
\label{def_level2}
When the protocol ends, \emph{Bob} learns the number of common attributes and the similarity value only when it exceeds the pre-defined threshold, \emph{Alice} learns the number of common attributes as well as the similarity value.
\end{definition}
This case indicates two threats from both sides. The illegally input on the \emph{initiator} side and curiously adjust the threshold on the \emph{responder} side.

\subsection{Cryptography Tools}
\label{subsec_tools}
Existing private matching solutions always rely on Private Set Intersection (PSI) \cite{PSI04, Sang09, Ateniese11} or Private Cardinality of Set Intersection (PCSI) \cite{Cristofaro10}, which is deeply based on heavy cryptographic operations such as Secure Multi-party Computation (SMC) \cite{SMC82}. To avoid the heavy cryptographic operations, our protocols utilize the commutative encryption function \cite{Agrawal03}, which is more computational friendly and satisfies the condition: $E_{k_1}(E_{k_2}(x))=E_{k_2}(E_{k_1}(x))$. Thus, a user who has the key $k_1$ or $k_2$ learns $x_1=x_2$ iff. $E_{k_1}(E_{k_2}(x_1))=E_{k_2}(E_{k_1}(x_2))$, but cannot learn any other $x_i$s of other user if $x_i$ is not a common attribute. Since the priority for every attribute is considered here, it is required that the encryption function needs to be easily deciphered to compute the similarity of two users. So we adopt the power function $f_{k}(x)=x^{k}\mod p$ as our encryption function for a safe prime $p$, i.e. $p$ and $(p-1)/2$ are both prime numbers. For all integers $k_1$, $k_2$ and $x\in \mathbb{Z}^*_p$, $\exists$ an integer $n$, s.t.
\begin{small}
\begin{eqnarray}
f_{k_1}(f_{k_2}(x)) & = & f_{k_1}(x^{k_2} \mod p) \nonumber\\
& = & f_{k_1}(x^{k_2}-np) \nonumber\\
& = & (x^{k_2}-np)^{k_1}\mod p \nonumber\\
& = & x^{k_2k_1}\mod p,
\end{eqnarray}
\end{small}
the last equality follows from the binomial theorem. Similarly it holds $f_{k_2}(f_{k_1}(x))=x^{k_1k_2}\mod p$. Therefore $f_{k_1}(f_{k_2}(x))=f_{k_2}(f_{k_1}(x))$. To obtain the decryption function, we need a corresponding number $k'$ to every $k$. Choose $k'$ such that $k'k=1\mod \phi(p)$, where $\phi(p)$ is the Euler phi-function of $p$ and $\phi(p)=p-1$ since $p$ is a prime. We use the Extended Euclidean Algorithm to yield $k'$ and let $g_{k}(y):=y^{k}\mod p$ for $y\in \mathbb{Z}^*_p$. Then
\begin{small}
\begin{eqnarray}
g_{k'}(f_k(x)) & = & g_{k'}(x^k\mod p) \nonumber\\
& = & x^{k'k}\mod p \nonumber\\
& = & x^{n\phi(p)+1}\mod p \nonumber\\
& = & x.
\end{eqnarray}
\end{small}
The last equality holds because of the Euler Theorem. To guarantee that $x$ and $y$ are both in $\mathbb{Z}^*_p$, we need a cryptographic hash function $h_p(\cdot)$ which has the quadratic residues modulo $p$ as its range.

\section{The Proposed Schemes}
\label{sec_schemes}
In this section, we present a suit of priority-aware private matching schemes. The basic version, \emph{P-match}, satisfies Privacy Level I. As an improvement to achieve Privacy Level II, we propose the enhanced version \emph{P-match$^+$}. Finally, the efficient version, \emph{E-match}, improves the performance significantly by avoiding the heavy cryptographic tools such as commutative encryption function.

In our scenario, there may be several users in a particular area at a particular time period. Specifically, each user holds a set of messages $\{\langle x_i, a_i\rangle, K_A, k_A, x_i \in X\}$, where $X$ is a set of attributes, $a_i$ is the corresponding priority of $x_i$, $K_A$ and $k_A$ are two secret keys. Moreover, all the procedures are $\mod p$ in our schemes. We use the notations shown in Table \ref{Notations}.
\begin{table}[!t]
\scriptsize
\renewcommand{\arraystretch}{1.5}
\caption{Notations}
\label{Notations}
\centering
\begin{tabular}{c||c}
\hline
$|A|$ &  Number of elements in the set $A$ \\
\hline
$R$ & Public attribute pool for all users, $n=|R|$, $R=\{r_i\}_{i=1}^{n}$ \\
\hline
$X$ & Attribute set of \emph{Alice}, $n_1=|X|,$ $X=\{x_i\}_{i=1}^{n_1}$, $x_i \in R$   \\
\hline
$Y$ & Attribute set of \emph{Bob}, $n_2=|Y|,$ $Y=\{y_i\}_{i=1}^{n_2}$, $y_i \in R$ \\ \hline
$S$ & $S=X \cap Y$, $q=|S|$  \\ \hline
$a_i(b_i)$ & Priority of $x_i(y_i)$, $a_i,\,b_i=\{1,2, \ldots, 10\}$  \\  \hline
$V_A$ & $V_A=\{a_i\}_{i=1}^q$, where each $a_i$ is the priority of $x_i \in S$  \\ \hline
$V_B$ & $V_B=\{b_i\}_{i=1}^q$, where each $b_i$ is the priority of $y_i \in S$ \\
\hline
$\xi^*$ & the expected value of the random variable $\xi$ \\
\hline
\end{tabular}
\end{table}

\subsection{Basic Version}
\label{subsec_pmatch}
\subsubsection{Introducing the basic similarity function}
Let \emph{Alice}'s attribute set and corresponding priority vector be $X$ and $A$ respectively, and \emph{Bob}'s attribute set and corresponding priority vector be $Y$ and $B$ respectively. The set of common attributes between \emph{Alice} and \emph{Bob} is denoted as $S=X \cap Y$, where $q=|S|$ and $S=\{s_1,s_2,\ldots,s_q\}$. Then we arrange the corresponding priorities of \emph{Alice} and \emph{Bob} on the common attributes into vector $V_A=(a_1, a_2,\ldots,a_q)$ and $V_B=(b_1, b_2, \ldots, b_q)$, respectively. The most widely applied similarity function is cosine similarity:
\begin{small}
\begin{eqnarray}
      \label{for_consine}
      \cos(\theta)=\frac{V_A\cdot V_B}{\|V_A\| \cdot \|V_B\|},
\end{eqnarray}
\end{small}
where $\theta$ is the angle between $V_A$ and $V_B$. It is often used to measure the angular similarity between two vectors. However, cosine similarity is orthogonal to the priorities on a common attribute. That is cosine similarity can be high when the priorities on the common attribute are quite different or exactly the same. This implies its defect. We do not accept the high similarity if the priorities differs far from each other. Thus, it is not the best choice in our scenario.

Different from cosine similarity, Jaccard similarity coefficient better fits our scenario, which considers the quotient of the size of the intersection over the size of the union of $X$ and $Y$. To simplify the computation we employ its variant form, Tanimoto similarity coefficient \cite{Ochiai57}:
\begin{small}
\begin{eqnarray}
      \label{for_Tanimoto}
      T(A,\,B)=\frac{V_A \cdot V_B}{\|V_A\|^2+\|V_B\|^2-V_A \cdot V_B},
\end{eqnarray}
\end{small}
where $V_A$ and $V_B$ are the same as in \eqref{for_consine}. The inner product appears in the numerator and the denominator of \eqref{for_Tanimoto} displays the difference between $V_A$ and $V_B$, and the norm term adjusts the size of the unit. After these refinement, Tanimoto similarity coefficient can embody the difference between two priority set on common attributes.

\subsubsection{P-match}
Based on the commutative encryption function \cite{Agrawal03} and Tanimoto similarity coefficient, we design our basic privacy-aware private matching scheme, \emph{P-match}, which considers both the number of common attributes and the corresponding priorities on them. As an initialization, users encrypt their attributes and priorities under their secret keys. Fig. \ref{fig_version1} shows the details and the procedure is as follows:
\begin{figure*}[!t]
  \centering
    \includegraphics[width=6.5in]{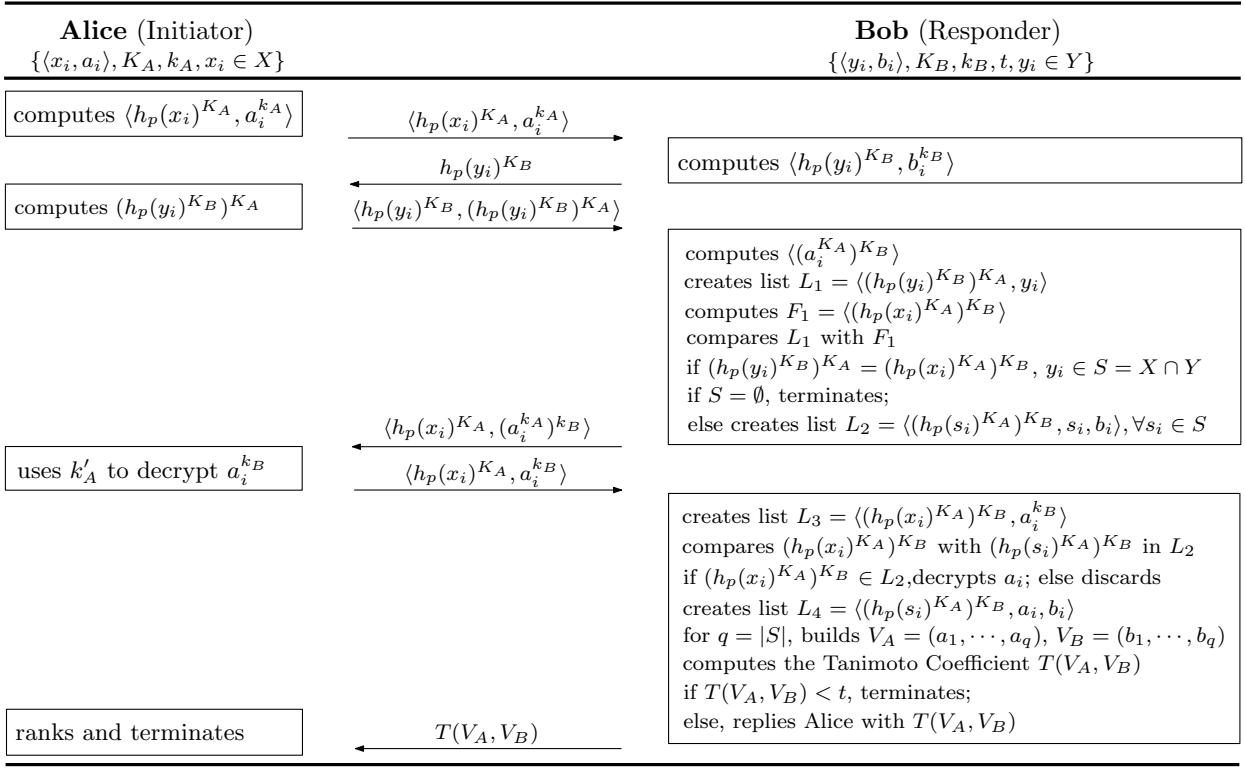}
  \caption{Basic Private Matching Scheme}
  \label{fig_version1}
\end{figure*}

\begin{enumerate}
  \item[(i)] When two users are within the communication range of each other, the \emph{initiator} \emph{Alice} begins the matching process by broadcasting a message $\langle h_p(x_i)^{K_A}, a_i^{k_A}\rangle$;
  \item[(ii)] as one of the \emph{responder}s, \emph{Bob} replies \emph{Alice} with $h_p(y_i)^{K_B}$;
  \item[(iii)] based the received message, \emph{Alice} computes $(h_p(y_i)^{K_B})^{K_A}$ and sends the two-tuple $\langle h_p(y_i)^{K_B}, (h_p(y_i)^{K_B})^{K_A}\rangle$ together to \emph{Bob};
  \item[(iv)] \emph{Bob} finds matches and creates a list $L_1 = \langle (h_p(y_i)^{K_B})^{K_A}, y_i\rangle$, and computes a set $F_1 = \langle (h_p(x_i)^{K_A})^{K_B}\rangle$. Then, he compares the first element in $L_1$ with $F_1$ to compute the common attributes with \emph{Alice}, the common ones form as set $S$, i.e., $S= X \cap Y$. \emph{Bob} creates another list $L_2 = \langle (h_p(s_i)^{K_A})^{K_B}, s_i, b_i\rangle, \forall s_i \in S$ unless $S = \emptyset$, and sends back the $\langle h_p(x_i)^{K_A}, (a_i^{k_A})^{k_B}\rangle$;
  \item[(v)] \emph{Alice} computes her $k_A'$ by Extended Euclidean algorithm and decrypts the second part of the received message, then sends out the results $\langle h_p(x_i)^{K_A}, a_i^{k_B}\rangle$;
  \item[(vi)] upon the received messages, \emph{Bob} creates another list $L_3 = \langle (h_p(x_i)^{K_A})^{K_B}, a_i^{k_B}\rangle$, then compares the first element in $L_3$ with $(h_p(s_i)^{K_A})^{K_B}$ in $L_2$. He obtains the corresponding $a_i$ by comparing if the received $a_i^{k_B} = b_i^{k_B}$ when he can find $(h_p(x_i)^{K_A})^{K_B} \in L_2$, otherwise, discards them. When he gets these information, another list is created as $L_4 = \langle (h_p(s_i)^{K_A})^{K_B}, a_i, b_i\rangle$. Then for $q = |S|$, \emph{Bob} uses priorities on each common attribute of \emph{Alice} and himself to build two vectors $V_A = (a_1, a_2, \cdots, a_q)$ and $V_B = (b_1, b_2, \cdots, b_q)$, respectively. To measure the similarity, he computes the Tanimoto Coefficient $T(V_A,V_B) = \frac{V_A \cdot V_B}{||V_A||^2 + ||V_B||^2 - V_A \cdot V_B}$, and compares with the threshold $t$ which is predefined by \emph{Bob}. If $T(V_A,V_B) < t$, the process terminates, otherwise, \emph{Bob} replies \emph{Alice} with $T(V_A,V_B)$.
\end{enumerate}


As the \emph{initiator} may receive several replies from others, a ranking of Tanimoto Coefficient is provided for the \emph{initiator} to choose the best match.

By utilizing \emph{P-match}, we achieve Privacy Level I. \emph{Bob} can learn the common attributes with \emph{Alice}, as well as the corresponding priorities, while \emph{Alice} learns nothing except the similarity value. However, there may be a problem when the scenario changes, i.e., \emph{Alice} wants to know some knowledge since \emph{Bob} may cheat on her. Thus, we propose an enhanced version \emph{P-match$^+$}.

\subsection{Enhanced Version}
\subsubsection{Constructing the enhanced similarity function}
When we only consider the common attributes and their priorities, Tanimoto similarity coefficient is energetic, but not effective enough if all the attributes and priorities are taken into account, because it is impossible to invent an efficient way to put \emph{Alice}'s and \emph{Bob}'s all attributes in the same order if $X\neq Y$. Hence we need another function, Ochiai similarity coefficient \cite{Lipkus99}
\begin{small}
\begin{equation}
O(A,\,B)=\frac{\,|A\cap B|}{\sqrt{|A|\cdot|B|}}.
\end{equation}
\end{small}
This coefficient was firstly applied in biology and is useful in comparing the faunistic feature between two different localities. It is also a ratio, where the numerator is the size of the intersection of two sets $A$, $B$, and the denominator is the geometric average between the size of $A$ and $B$. However, since all the priorities, on common and uncommon attributes, are considered here, and we cannot do the intersection between two ordered priority sets, we need some more effective similarity function. For that purpose, we notice that our priority is a kind of weight function. For a finite set $Z=\{z_1, z_2, \ldots, z_{\lambda}\}$ where every $z_{i}$ has the weight $w(z_i)$, $i=1,2,\ldots, \lambda$, let the weight function be $w: Z \rightarrow \mathbb{I}$, then the un-weighted sum on $Z$ is defined by $\sum^{\lambda}_{i=1}z_i$ while the weighted sum on $Z$ is defined by $\sum^{\lambda}_{i=1}{z_i w(z_i)}$. If we let $Z=X$, $\mathbb{I}=\{1,\ldots,9\}$ and $f(x_i)=a_i$, then the weighted model becomes our scenario. This implies that we can regard the priority as a counting rule of an attribute, which is, we count an attribute $x_i$ $a_i$ times.

Therefore, we construct a new priority-aware coefficient $P(A,\,B)$ based on Ochiai similarity coefficient and the weighted sample. For two attribute sets $X=\{x_1,x_2,\ldots,x_{n_1}\}$ and $Y=\{y_1,y_2,\ldots,y_{n_2}\}$, the corresponding priority vectors are $A=(a_1, a_2, \cdots,a_{n_1})$ and $B=(b_1, b_2, \cdots, b_{n_2})$, let $S=X \cap Y=\{s_1, \ldots,s_q\}$. First we generate two counting sets:
\begin{small}
\begin{eqnarray} \label{X prime}
X'&=&\{x_1,x_1+1,\ldots,x_1+a_1, \\ \nonumber
&& x_2,x_2+1,\ldots,x_2+a_2, \ \ldots\\ \nonumber
&& x_{n_1},x_{n_1}+1,\ldots,x_{n_1}+a_{n_1}\},
\end{eqnarray}
\end{small}
\begin{small}
\begin{eqnarray} \label{Y prime}
Y'&=&\{y_1,y_1+1,\ldots,y_1+b_1, \\ \nonumber
&& y_2,y_2+1,\ldots,y_2+b_2, \ \ldots\\ \nonumber
&& y_{n_2},y_{n_2}+1,\ldots,y_{n_2}+b_{n_2}\}.
\end{eqnarray}
\end{small}
Then \begin{eqnarray}\label{size prime} |X'|=\sum^{n_1}_{i=1}a_i,\ |Y'|=\sum^{n_2}_{i=1}b_i.\end{eqnarray}
Next we find the intersection of $X'$ and $Y'$,
\begin{small}
\begin{eqnarray}\label{intersection prime}
X' \cap Y'&=&\{s_1,s_1+1,\ldots,s_1+c_1, \\ \nonumber
&& s_2,s_2+1,\ldots,s_2+c_2, \ \ldots\\ \nonumber
&& s_q,s_q+1,\ldots,s_q+c_q\},
\end{eqnarray}
\end{small}
where every $c_i=\min\{a_i,b_i\}$ for $i=1,\ldots,q$.
Then \begin{small}\begin{eqnarray}\label{intersection size prime} |X' \cap Y'|=\sum^{q}_{i=1}c_i=\sum^{q}_{i=1}\min\{a_i,b_i\}.\end{eqnarray}\end{small}
Provided \eqref{size prime} and \eqref{intersection size prime}, now we apply the Ochiai similarity coefficient to the two counting sets $X'$ and $Y'$ given by \eqref{X prime} and \eqref{Y prime}, and define it to be our new similarity function:
\begin{small}
\begin{eqnarray}
\label{for_weighted}
      P(A,\,B) & = & O(X',\,Y') \nonumber\\
               & = & \frac{\;\left|X'\cap Y'\right|}{\sqrt{|X'| \cdot |Y'|}} \nonumber\\
               & = & \frac{\overset{q}{\underset{i=1}{\sum}} \min \{a_i,\,b_i\}}{\sqrt{ \sum_{i=1}^{n_1} a_i} \cdot \sqrt{ \sum_{i=1}^{n_2} b_i}}.
      \end{eqnarray}
      \end{small}
The range of the priority-aware similarity coefficient \eqref{for_weighted} is from $0$ to $1$, which are corresponding to the cases "no common attribute at all" for 0 and "same attributes, same priorities" for 1.


\subsubsection{P-match$^+$}
To achieve Privacy Level II, we change some procedures from step vi in our basic version.
\begin{algorithm}
\caption{Part of Enhanced Version on Responder Side}
\label{alg_enhanced_responder}
\SetKwInOut{Input}{Input}
\SetKwInOut{Output}{Output}
\Input{when receives $\langle h_p(x_i)^{K_A}, a_i^{k_B}\rangle$ from \emph{Alice}}
\Output{$n, P(A,B)$ to \emph{Alice}}
computes $(h_p(x_i)^{K_A})^{K_B}$\;
sends $\langle h_p(x_i)^{K_A}, (h_p(x_i)^{K_A})^{K_B}\rangle$ to \emph{Alice}\;
computes $k_B'$ to decrypt $a_i$\;
creates $L_3 = \langle (h_p(x_i)^{K_A})^{K_B}, a_i\rangle$\;
compares $(h_p(x_i)^{K_A})^{K_B}$ in $L_3$ with $(h_p(s_i)^{K_A})^{K_B}$ in $L_2$\;
for those $(h_p(x_i)^{K_A})^{K_B}$ in $L_2$, computes $c_i = \min \{a_i, b_i\}$\;
$P(A, B) = \frac {\sum_{i=1}^{q} c_i}{\sqrt{\sum_{i=1}^{|X|} a_i} \cdot \sqrt{\sum_{i=1}^{|Y|} b_i}}$, where $q = |S|$\;
\uIf{$(P(A, B) < t)$}
    {
    terminates\;
    }
    \Else
    {
    replies \emph{Alice} with $P(A,B)$
    }
\end{algorithm}

Algorithm \ref{alg_enhanced_responder} shows the changes on \emph{responder} side in our \emph{P-match$^+$}. When \emph{Bob} receives $\langle h_p(x_i)^{K_A}, a_i^{k_B}\rangle$ from \emph{Alice}, he computes $(h_p(x_i)^{K_A})^{K_B}$ and sends them back to \emph{Alice} together with $h_p(x_i)^{K_A}$. He can also decrypt $a_i$ by computing $k_B'$. Then he creates a list $L_3 = \langle (h_p(x_i)^{K_A})^{K_B}, a_i\rangle$ and compares the first element in $L_3$ with $(h_p(s_i)^{K_A})^{K_B}$ in $L_2$. Followed, he computes $c_i = \min \{a_i, b_i\}$ for those $(h_p(x_i)^{K_A})^{K_B}$ in $L_2$, the priority-aware Ochiai Coefficient can be computed as $P(A, B) = \frac {\sum_{i=1}^{q} c_i}{\sqrt{\sum_{i=1}^{|X|} a_i} \cdot \sqrt{\sum_{i=1}^{|Y|} b_i}}$, where $q=|S|$. If $P(A, B) < t$, the algorithm terminates, otherwise, \emph{Bob} replies $P(A,B)$ to \emph{Alice}.

\begin{algorithm}
\caption{Part of Enhanced Version on Initiator Side}
\label{alg_enhanced_initiator}
\SetKwInOut{Input}{Input}
\SetKwInOut{Output}{Output}
\Input{when receives $\langle h_p(x_i)^{K_A}, (h_p(x_i)^{K_A})^{K_B}\rangle$ from \emph{Bob}}
\Output{The number of common attributes with \emph{Bob}}
creates $\langle x_i, (h_p(x_i)^{K_A})^{K_B}\rangle$\;
compares $(h_p(x_i)^{K_A})^{K_B}$ with $(h_p(y_i)^{K_B})^{K_A}$\;
\uIf{$((h_p(x_i)^{K_A})^{K_B} = (h_p(y_i)^{K_B})^{K_A})$}
    {
    $s_i = x_i$ and $S=S \cup \{s_i\}$\;
    }
    \Else
    {
    terminates\;
    }
outputs $|S|$\;
\end{algorithm}
While on the \emph{initiator} side, \emph{Alice} needs to do some computation work to obtain the number of the common attributes, the details are shown in Algorithm \ref{alg_enhanced_initiator}. Based on the received message $\langle h_p(x_i)^{K_A}, (h_p(x_i)^{K_A})^{K_B}\rangle$ from \emph{Bob}, the algorithm creates a list of $\langle x_i, (h_p(x_i)^{K_A})^{K_B}\rangle$, and obtains the set of common attributes by comparing $(h_p(x_i)^{K_A})^{K_B}$ in this list with $(h_p(y_i)^{K_B})^{K_A}$, which is received before. It terminates if there is no match found by computing $(h_p(x_i)^{K_A})^{K_B} = (h_p(y_i)^{K_B})^{K_A}$, otherwise, outputs $|S|$ to \emph{Alice}.

\begin{figure*}[!t]
  \centering
    \includegraphics[width=6.5in]{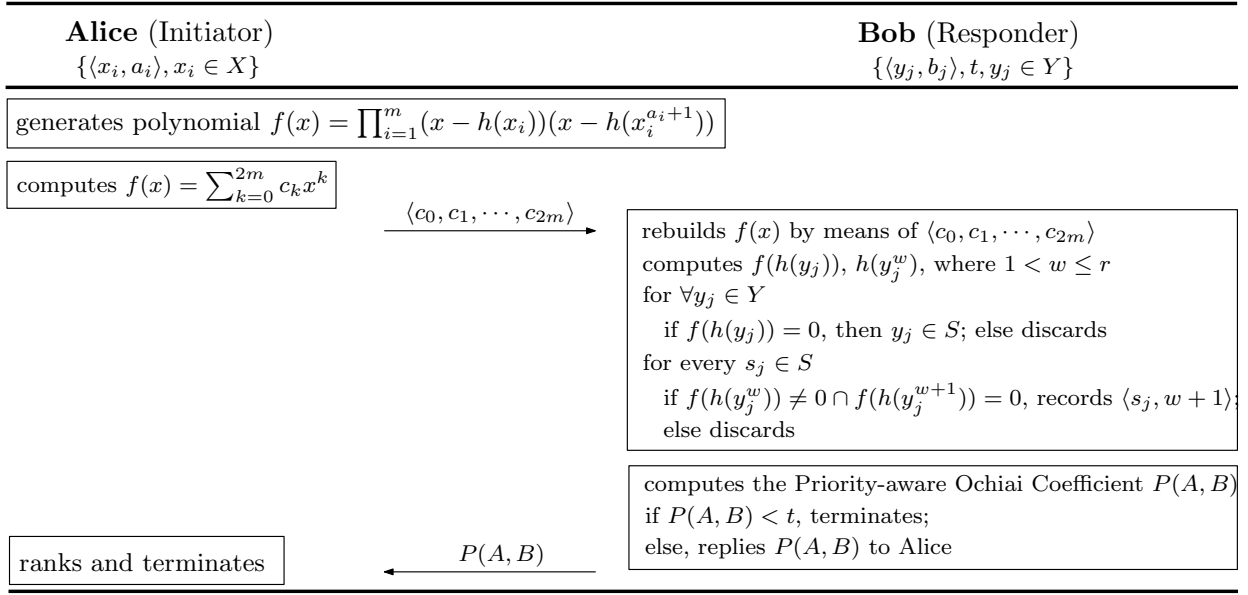}
  \caption{Efficient Private Matching Scheme}
  \label{fig_version2}
\end{figure*}

\subsection{Efficient Version}
For both \emph{P-match} and \emph{P-match}$^+$, they can achieve their designed goals at the cost of system performance. Since the heavy cryptographic operations, such as the keyed hash functions and exponentiation operations, are hard to perform in current mobile devices, the performance drops dramatically when the number of attributes goes large. We thus design a more efficient version, \emph{E-match}, which employs a Bloom filter \cite{Bloom70} instead of the heavy exponentiation operations in an honest-but-curious environment. A Bloom filter is a space-efficient probabilistic data structure which is used to test whether an element is a member of a set.

\subsubsection{Initialization}
Suppose that the public database is $R=\{r_i\}^{n}_{i=1}$ consisting of attributes of all users. We let $\{h_i(\cdot)\}_{i=1}^l$ be a family of hash functions with $h_j(r_{i}) \in [1, \lambda]$ for an attribute $r_{i} \in R$, and $\mathcal{H}$ be a large public pool of hash functions of such $h_i$ with each indexed by a unique identifier. An empty $\lambda$-bit Bloom filter is an array of $\lambda$ bits, all setting to $0$ bits. To add an element $r_{i}$ to the $\lambda$-bit Bloom filter, we set all the bits in $h_j(r_i)$ positions $1$, $1 \leq j \leq l$. To check whether an $r_i$ is some user's attribute, we verify whether all the bits in positions $h_j(r_i)$ are $1$. If not, $r_i$ is not this user's attribute; otherwise, $r_i$ is his/her attribute with a probability determined by $n, \lambda$ and $l$.

We label each $r_i \in R$ by the 2-tuple $\{j,r_i(j)\}^{\kappa}_{j=1}$, where $r_i(j)=r_i+j-1$ is the counting function of $r_i$. Then the database $R$ is extended to the set $\{i,\{j,r_i(j)\}^{\kappa}_{j=1}\}^{n}_{i=1}$. Moreover, this extended set can be identified with the indexed set $R'=\{\{i,j,r_i(j)\}^{\kappa}_{j=1}\}^{n}_{i=1}$. If \emph{Alice} has an attribute set $X=\{x_i\}_{i=1}^{n_1}$ with the priority set $\{a_i\}^{n_1}_{i=1}$, we can assign her a personal set $S_A=\{\{i,j,x_i(j)\}^{a_i}_{j=1}\}^{n_1}_{i=1}$, where $x_i=r_{i'}$ for an attribute $r_{i'} \in R$, and $a_i \in \{1,2, \cdots, \kappa\}$ is the priority of $x_i$. The same technique can be applied to \emph{Bob}, we get $S_B=\{\{i,j,y_i(j)\}^{b_i}_{j=1}\}^{n_2}_{i=1}$, where $y_i$ is in \emph{Bob}'s attribute set $Y$, $|Y|=n_2$, $b_i$ is the priority of $y_i$. Denote $q_1:= |S_A|$, $q_2:=|S_B|$, then we have $q_1=\Sigma_{i=1}^{n_1} a_i = |X'|$, $q_2=\Sigma_{i=1}^{n_2} b_i = |Y'|$ with $X'$ given by Equation \ref{X prime}, and $Y'$ given by Equation \ref{Y prime}.

We pair every $l$ a random number $l'$, $1<l'<l$, and publish the $2$-tuple $(l,l')$ to all users. Now we can use a $\lambda$-bit Bloom filter to check how many attributes are in \emph{Alice}' set $X$ with her priorities, and how many are in \emph{Bob}'s set $Y$ with his priorities, so that one of them can learn the similarity level.

\subsubsection{E-match}
We then present our proposed \emph{E-match} shown in Fig. \ref{fig_version2} in details.
\begin{enumerate}
  \item[(i)] \emph{Alice} sends a request to \emph{Bob}.
  \item[(ii)] \emph{Bob} agrees.
  \item[(iii)] \emph{Alice} randomly chooses $\{h_i\}_{i=1}^{l}\subset\mathcal{H}$ with indexes denoted by $\mathcal{H}_A$. \emph{Alice} adds each $\{i,j,x_i(j)\}$ in $S_A$ into a $\lambda$-bit Bloom filter array, denoted by $BF_{A}$, with $l'$ different random hash functions in $\mathcal{H}_A$ and $(l-l')$ random hash functions out of $\mathcal{H}$ (Computation offline above). \emph{Alice} sends $\mathcal{H}_A$ and $BF_A$ to \emph{Bob}.
  \item[(iv)] \emph{Bob} counts the number of $0$-bits in $BF_A$, denoted by $d_1$, then he adds every $\{i,j,y_i(j)\}$ in $S_B$ to $BF_{A}$ using $\mathcal{H}_A$ to get a $\lambda$-bit Bloom filter array, denoted by $BF_B$. He counts the number of $0$ bits $BF_B$, say, $d_0$, and computes
      \begin{small}\be \label{p star}
      P^*(A,B)=\frac{\sqrt{l}[q_2-\lambda(\ln d_1 -\ln d_0)]}{l' \sqrt{\lambda q_2(\ln \lambda -\ln d_1)}},
      \ee
      \end{small}
      \emph{Bob} compares $P^*(A,B)$ with his pre-defined threshold $t$, so that he decides whether to match \emph{Alice} or not.
\end{enumerate}

\section{Security Analysis}
\label{sec_secur}
In this section, we prove that our proposed schemes can achieve the required privacy level in turns.

\subsection{Analysis of the Basic Scheme}
\label{subsec_analysis_basic}

\begin{theorem}
\label{the_basic}
P-match ensures Privacy Level I if the commutative encryption function is secure.
\end{theorem}
\begin{myproof}
\textbf{For \emph{Alice.}} $Bob$ encrypts the hash value of his attributes using his secret keys $K_B$ and $k_B$, then sends the messages ($h_p(y_i)^{K_B}$ and $\langle h_p(x_i)^{K_A}, (a_i^{k_A})^{k_B}\rangle$) to \emph{Alice} side. As mentioned in Section 3.4, the commutative encryption function is secure, so it is computationally impossible to \emph{Alice} to obtain any of the attributes or the corresponding priorities of $Bob$. When the protocol ends, what \emph{Alice} is able to get is a similarity value, which, in general, indicates a rough similarity. However, it is impossible to deduce any personal information, such as the number of common attributes or the corresponding priorities, from the similarity value.

\textbf{For \emph{Bob.}} As a \emph{responder}, \emph{Bob} learns more information than \emph{Alice}. It is reasonable because a \emph{responder} has the weaker motivation to start an attack. In \emph{P-match}, even if this kind of attacks happens, the commutative encryption function and cryptographic hash function can guarantee that what \emph{Bob} gets are only the common attributes and the corresponding priorities of \emph{Alice}, since all encryption function are injective, i.e.  $(h_p(x_i)^{K_A})^{K_B} = (h_p(y_i)^{K_B})^{K_A}$ if and only if $x_i = y_i$, it holds the similar conclusion for the corresponding priorities. So, \emph{Bob} knows nothing about \emph{Alice} except the common attributes and priorities so that he can compute the similarity value on his own side.
\end{myproof}

\subsection{Analysis of the Enhanced Version}
\label{subsec_analysis_enhanced}
\begin{theorem}
\label{the_enhance}
P-match$^+$ ensures Privacy Level II if the commutative encryption function is secure.
\end{theorem}
\begin{myproof}
Similar to the proof of Theorem \ref{the_basic}, the commutative encryption function and keyed hash function provide end users with a secure channel, it means that only the one who has the secret key can decrypt the message if he has at least one common attribute with the other user. Here, the following two possible threats are our focal points in this match, 1) \emph{Alice} may illegally input her attributes as well as the priorities; 2) \emph{Bob} tries to learn extra information from the received messages from \emph{Alice} by adjusting his threshold.

\textbf{For case 1)}, the more \emph{Alice} inputs attributes $x_i$ with the higher priorities $a_i$, the more personal information of \emph{Bob} she gets easily and possibly. We note that for the priority-aware Ochiai coefficient, the denominator of (\ref{for_weighted}) has the factor $\sqrt{\sum_{i=1}^{|X|}a_i}$, which goes large quickly if either $|X|$ or any $|a_i|$ becomes large. Meanwhile, the numerator $\sum_{i=1}^{q} \min \{a_i,\,b_i\}$ and the other term $\sqrt{\sum_{i=1}^{|Y|}b_i}$ in the denominator are preserved for the same responder \emph{Bob}. Hence, if \emph{Alice} is a malicious user, the priority-aware Ochiai coefficient between her and any other responder \emph{Bob} decreases dramatically so that \emph{Bob} terminates the protocol according to his pre-defined threshold, and \emph{Alice}'s attack fails.

\textbf{For case 2)}, to get more extra information from \emph{Alice}, \emph{Bob} does not want to terminate the protocol, so he must lower his threshold $t$ to proceed Algorithm \ref{alg_enhanced_responder}. But in P-match$^+$ \emph{Alice} does not send \emph{Bob} anything except $h_p(x_i)^{K_A}$, $a_i^{k_A}$ and $a_i^{k_B}$ in the beginning phase. Since the encryption function $f_k(x)=x^k$ is pre-image resistant, so \emph{Alice}' information which is known to \emph{Bob} is not change at all, whether \emph{Bob} adjusts his threshold. That means \emph{Bob} cannot get more information from \emph{Alice} by lowering the threshold $t$.

Considering the proof of Theorem \ref{the_basic}, we obtain the security for the enhanced version P-match$^+$.
\end{myproof}

\subsection{Analysis of the Priorities}
Agrawal \emph{et al}. \cite{Agrawal03} proved that, it computes impossible to map $x_i^k$ to $x_i$ without knowing the secret key $k$ under the Decisional Diffie-Hellman hypothesis (DDH). Specifically, for fixed values of $i$ and $j$, $\langle x_i, f_k(x_i), y_j, f_k(y_j) \rangle$ is indistinguishable from $\langle x_i, f_k(x_i), y_j, z \rangle$, where $f_k(x)=x^k \mod p$. That means, \emph{Bob} cannot map $x_i^k$ back to $x_i$ if he does not know the value of $k$, which is \emph{Alice}'s secret key and only known to \emph{Alice}. Based on these conclusions, each user's attributes are safe. Now we consider the security on the priorities. For the two similarity functions $T(A, B)$ and $P(A, B)$, note that \emph{Alice} knows the common attributes and \emph{Bob}'s corresponding priorities if $T(A, B)=1$ and $|X|=|X\cap Y|$ in the basic version, and if $P(A, B)=1$ in the enhanced version. Otherwise, when the two conditions do not hold, we have the following results.
\begin{theorem}
\label{the_basic_one}
In P-match, if Alice has only one attribute, she can confirm a common attribute and the corresponding priority of Bob iff. there is only one common attribute between them. Otherwise, Alice knows nothing if she has more than one attribute.
\end{theorem}
\begin{myproof}
When \emph{Alice} and \emph{Bob} have only one common attribute with priorities $a$ and $b$, respectively, based on \emph{P-match}, she can easily compute $b$ from $T(A, B)=\frac{a \cdot b}{a^2 + b^2 - a \cdot b}$, where $a$ and $T(A, B)$ are known to \emph{Alice}. The other direction is trivial. Now we assume \emph{Alice} has more than one attribute. Since in \emph{P-match} we only take the common attributes and their priorities into account, and we have proved that \emph{P-match} ensures Privacy Level I if the commutative encryption function is secure in Theorem \ref{the_basic}, which means  \emph{Alice} knows nothing if the encryption function is secure. We note that the attributes and corresponding priorities are encrypted by two different secret keys, and the security of the attribute part was proved by Agrawal \emph{et al}. in \cite{Agrawal03}, then \emph{Alice} knows nothing.
\end{myproof}

\begin{theorem}
\label{the_enhance_one}
In P-match$^+$, Alice can confirm a common attribute as well as the corresponding priority of Bob iff. she has only one attribute which is also the only one common attribute. Otherwise, Alice would know at most one common attribute and its corresponding priority of Bob.
\end{theorem}
\begin{myproof}
The proof for the first part is similar to Theorem \ref{the_basic_one}. Since in this enhanced version, Privacy Level II can be ensured if the commutative encryption function is secure, which means, \emph{Alice} knows the common attributes. If \emph{Alice} has at least two attributes and they have the common attributes of number $q \geq 2$, from the expression of $P(A, B)$, \emph{Alice} only knows the ratio $\frac{\sum _{i=1}^q \min\{a_i,b_i\}} {\sqrt{{\sum_{i=1}^m}b_i}}$. However, this ratio means nothing if \emph{Alice} does not know the right priority of every common attribute.
\end{myproof}

As stated above, we see that there is other potential attack that a malicious user inputs only one attribute with the corresponding priority to learn other's secret information. However, there are many practical ways to figure out this problem, such as setting a rule to limit the minimum number of input. Thus, it is omitted here.

\subsection{Analysis of the E-match}
The security of the \emph{E-match} is based on the probability of false positives for the $\lambda$-bit Bloom filter. The accuracy of the \emph{E-match} is given by the following two results.
\begin{theorem}
\label{accuracy}
Via our protocol, $Bob$ can know the expected values of counts $q_1=|S_A|$ and $q':=|S_A \cap S_B|$, by
\begin{small}
\be \label{q1 star}
q_1^*=\frac{\lambda (\ln \lambda -\ln d_1)}{l},
\ee
\end{small}
and
\begin{small}
\be \label{q' star}
q'^*=\frac{l q_2+\lambda (\ln d_0 -\ln d_1)}{l'},
\ee
\end{small}
where $d_1$ is the number of $0$-bits in $BF_A$, $d_0$ is the number of $0$-bits in $BF_B$.
\end{theorem}
\begin{myproof}
According to \cite{CHE07}, the distribution for $0$-bits in a $\lambda$-Bloom filter can be regarded as a binomial distribution. When the length $\lambda$ is large enough, it approximates a normal distribution asymptotically. We now suppose that $\lambda$ is large enough. For a fixed bit in $BF_A$, the probability that it is set to $0$ by adding one element $x_i(j)$ with $l$ hash functions is $(1-\frac{1}{\lambda})^l$, the probability that it is set to $0$ by adding all elements $x_i(j)$ with $l$ hash functions is
$$(1-\frac{1}{\lambda})^{l q_1}\thickapprox e^{-\frac{l q_1}{\lambda}}.$$ Thus we have $e^{-\frac{l q_1}{\lambda}}=\frac{d_1}{\lambda}$. Solving this equation leads to $q_1^*$ given by the equation\eqref{q1 star}.

For a fixed bit in $BF_A$, the probability that it is set to $0$ by adding $q'$ common element with $l'$ hash functions is $(1-\frac{1}{\lambda})^{l'q'}$, the probability that it is set to $0$ not by adding $q'$ common element with $l'$ hash functions is $(1-\frac{1}{\lambda})^{lq_1-l'q'}$. Thus for a bit in $BF_B$, the conditional probability that it is set to $0$ is $$(1-\frac{1}{\lambda})^{lq_2+l q_1-l'q'} \thickapprox e^{-\frac{l q_1+l q_2 -l'q'}{\lambda}}.$$ Then we have $$e^{-\frac{l q_1+l q_2 -l'q'}{\lambda}}=\frac{d_0}{\lambda}.$$ Solving this equation, and combining with the Equation \eqref{q1 star} lead to $q'^*$ in the Equation \eqref{q' star}. If we note that the Equation \eqref{for_weighted} of Ochiai coefficient implies that $$P(A,B)=\frac{q'}{\sqrt {q_1 q_2}},$$ The Equation \eqref{p star} is obtained from the Equation \eqref{q1 star} and Equation \eqref{q' star}.
\end{myproof}

\begin{theorem}
\label{distribution}
In the E-match, let $q_2=|S_A|$, then, with the same notations shown in Theorem \ref{accuracy}, we have
\begin{small}
\be \label{q1 star distribution}
q_1^* \sim \mcn [q_1, \frac{\lambda}{l^2} (e^{\frac{lq_1}{\lambda}}-1)],
\ee
\end{small}
and
\begin{small}
\be \label{q' star distribution}
q'^* \sim \mcn [q', \frac{\lambda}{(l')^2} (e^{\frac{lq_1}{\lambda}}+e^{\zeta}-2-\zeta)]
\ee
\end{small}
with $\zeta=\frac{lq_1+lq_2 -l'q'}{\lam}$.
\end{theorem}
\begin{myproof} From \cite{CHE07} we know that when $\lam$ is large enough,
\begin{small}
\be \label{d1 distribution}
d_1 \sim \mcn [\mu_1(q_1), \sig_1^2 (q_1)]
\ee
\end{small}
where
\begin{small}
\ben
\mu_1(q_1)=\lam (1-\frac{1}{\lam})^{lq_1}\approx \lam e^{-\frac{lq_1}{\lam}},
\een
\end{small}
\begin{small}
\ben
\sig_1^2(q_1)=  \lam (1-\frac{1}{\lam})^{lq_1}[1-(1-\frac{1}{\lam})^{lq_1}] \approx\lam e^{-\frac{lq_1}{\lam}} (1-e^{-\frac{lq_1}{\lam}}).
\een
\end{small}
To apply the special version of the central limit theorem stated in the Theorem 6 in \cite{Kodialam06}, we let $\lam \rightarrow \infty$, $lq_1 \rightarrow \infty$ while $\frac{lq_1}{\lam}$ is fixed. Regard $q_1$ as a variable, and note that $\mu_1(q_1)$ is monotonically decreasing, then $\mu_1$ has an inverse function, denoted by $g_1$. Based on the Theorem 6 in \cite{Kodialam06} and the Equation \eqref{d1 distribution}, we know that
\begin{small}
\be\label{g1d1}
g_1(d_1) \sim \mcn [g_1(\mu_1(q_1)), \delta_1^2(q_1)],
\ee
\end{small}
where $\delta_1^2(q_1)=\sig_1^2 (q_1)(g_1 '(\mu_1(q_1)) )^2$. Since $g_1(d_1)=q_1^*$, and
\begin{small}
\be
g_1 '(\mu_1(q_1))=\frac{1}{\mu'(q_1)}=-l e^{-\frac{l}{q_1}},
\ee
\end{small}
we thus obtain the result in Equation \eqref{q1 star distribution}.

To prove \eqref{q' star distribution}, we let
\begin{small}
\be \label{q'0 star}
q'^*_0=\frac{lq_1+lq_2-\lam (\ln \lam - \ln d_0)}{l'}.
\ee
\end{small}
Then the same technique as used in \cite{Sun13} results in
\begin{small}
\be \label{q'0 star distribution}
q'^*_0 \sim \mcn [q', \frac{\lam (e^{\zeta}-1-\zeta)}{(l')^2}].
\ee
\end{small}
Since the distributions of $q_1^*$ and $q'^*_0$ are both normal distributions, the distribution of the 2-tuple $(q_1^*, q'^*_0)$ is a multivariable normal distribution. Note that $q'^*=q'^*_0+\frac{l}{l'} q_1^*-\frac{l}{l'} q_1$, thus from the Equations \eqref{q1 star distribution}, and \eqref{q'0 star distribution} we have the Equation \eqref{q' star distribution}.\end{myproof}

The estimation of the \emph{E-match} is given by the following theorem.
\begin{theorem}
\label{accuracy estimation}
In E-match, let $q^*_1$ and $q'^*$ be defined by the Equations \eqref{q1 star} and \eqref{q' star}, respectively. Then for $\epsilon_1,\ \epsilon_2$ such that
$$\epsilon_1 q_1 \geq \frac{\left(\lambda (e^{\frac{lq_1}{\lambda}}-1)\right)^{\frac 1 2}}{l}$$ and $$\qquad   \epsilon_2 q' \geq \frac{\left(\lambda (e^{\frac{lq_1}{\lambda}}+e^{\zeta}-2-\zeta)\right)^{\frac 1 2}}{l'},$$ we have
\begin{footnotesize}
\be \label{estimation}
Pr(|q^*_1-q_1|\leq \epsilon_1 q_1) \geq 1- p_1, \quad Pr(|q'^*-q'|\leq \epsilon_2 q') \geq 1- p_2,
\ee
\end{footnotesize}
where
$$p_1 \geq \frac{\lambda (e^{\frac{lq_1}{\lambda}}-1)}{\epsilon_1^2 q^2_1 l^2}, \quad p_2 \geq \frac{\lambda (e^{\frac{lq_1}{\lambda}}+e^{\zeta}-2-\zeta)}{\epsilon^2_2(q'l')^2}.$$
\end{theorem}
\begin{myproof}
Combining the Equations \eqref{q1 star distribution}, \eqref{q' star distribution} with Chebyshev's inequality leads to the result in Equation \eqref{estimation}. \end{myproof}

We qualify the priority-aware attribute by the Shannon entropy, which is a common measurement of uncertainty. In the \emph{E-match}, only \emph{Alice} sends her information to other users, so that we only examine \emph{Alice}'s set $S_A$, where $S_A \subseteq R'$. On \emph{Bob}'s side, he only knows $\kappa \, n$ before the protocol, and he will know $\lam$, $l$ and $l'$ after the protocol. He will compute the expected value of $|S_A|$ but will not know any explicit attribute in $S_A$. Considering the size of $R'$ and $S_A$, there are $\frac{(\kappa n)!}{q_1 !\,(\kappa n-q_1)!}$ choices of $S_A \subseteq R'$. Among those choices, although some sets are counted repeatedly, $Bob$ does not know which sets are the same since he cannot know any of the priority $a_i$. That means, the $\frac{(\kappa n)!}{q_1 !\,(\kappa n-q_1)!}$ candidate sets are all equally unknown in $Bob$'s eyes. We replace $q_1$ by $q_1^*$.
Thus the uncertain attribute of \emph{Alice} to \emph{Bob} can be estimated in bits by
$$\e^*=\log_2 \frac{(\kappa n)!}{q_1 !\,(\kappa n-q_1)!}.   $$
Moreover we have the following result for the entropy.

\begin{theorem}
\label{Privacy analysis}
In the E-match, suppose that $BF_A$ is the Bloom filter constructed by $Alice$ using $l'$ hash functions in $\mathcal{H}_A$ based on her priority-aware personal set $S_A$. Then after sending $BF_A$ and $\mathcal{H}_A$ to Bob, her remaining privacy information of $S_A$ against Bob is
\begin{small}
\be \label{entropy estimation}
\e=q_1 \e[i,j],
\ee
\end{small}
where
\begin{small}
\ben
\displaystyle \e[i,j]&=&\sum^{\kappa n}_{x=1} {\kappa n \choose x}P^x (1-P)^{(\kappa n-x)} \log_2 x,\\
\displaystyle P&=&\sum^{l}_{i=1} {l \choose i} p^i (1-p)^{l-i},\\
\displaystyle p&=&1-e^{-\frac{lq_1}{\lam}}.
\een
\end{small}
\end{theorem}
\begin{myproof}
We refer to Theorem 2 in \cite{Sun13} for the privacy-preserving spatiotemporal matching. The result in Equation \eqref{entropy estimation} can be obtained if we consider each possible location cell (cID) to be each possible interest $r_i[j]$, and note that the size of the whole interest pool $R'$ is $\kappa n$.
\end{myproof}

\section{Performance Evaluations}
\label{sec_evalu}
In this section, we first design an experiment to verify the correctness of our proposed protocols. Then we analyze the system complexity and show our experimental results.
\subsection{Experiment of Correctness}
To verify our work, we design a simple experiment by letting the \emph{initiator} \emph{Alice} do the matching with several candidates (5 users in our experiment) in vicinity. For simplicity, we assign 5 attributes for each user, they can set the priorities on each individual from 1 to 9 randomly. As a result, each user may have dozens of combinations on these attributes. We choose a snapshot of these priorities, which is shown in Table \ref{tab_experiment_setting}. The notation "-" means having no interest at all. Then we compute the matching results of several existing work.

\begin{table}[!t]
\renewcommand{\arraystretch}{1.3}
\caption{Experiment setting}
\label{tab_experiment_setting}
\centering
\begin{tabular}{c|c|c|c|c|c}
\hline
& Cancer & Music & Football & Tennis & Cooking \\
\hline
\hline
\emph{Alice} & 8 & 4 & 1 & 3 & 2  \\
\emph{Bob} & 7 & - & 2 & - & -  \\
\emph{Charles} & 1 & 9 & 4 & 2 & 1  \\
\emph{David} & 9 & 8 & - & 6 & -  \\
\emph{Emmy} & - & 2 & 9 & 1 & 1  \\
\emph{Frank} & 8 & 3 & - & - & -  \\
\hline
\end{tabular}
\end{table}

The best match of \emph{Alice} in FindU \cite{FindU11} is \emph{Charles}, cause they have 5 common attributes. Algorithm in \cite{FineG12} considers the difference of priorities on each common attributes, as a result, it is hard to choose the best candidate from \emph{Bob} and \emph{David}, since the differences on these attributes are: \emph{Alice} to \emph{Bob}, $[1, 4, 1, 3, 2]$; and \emph{Alice} to \emph{David}, $[1, 4, 1, 3, 2]$. While in \emph{P-match}, the similarities with the nearby users are 0.9667, 0.3972, 0.8243, 0.2316, and 0.9870, respectively. Definitely, we prefer \emph{Frank} as the best match. Because either \emph{Alice} or \emph{Frank} is more interested in \emph{Movie} and \emph{Music} in their common attributes, even though they have only these two common attributes. We also compute the similarity values of \emph{P-match$^+$}, the results are 0.1122, 0.0905, 0.1138, 0.0547 and 0.1314, respectively. The best match is still \emph{Frank}. Similarly, the matching result in \emph{E-match} can also be computed by the priority-aware Ochiai similarity coefficient with a probability determined by $\lam$ and $l$, $l'$, then the matching results are same with \emph{P-match$^+$}. For instance, if we let $\lam=400$, $l=12$, $l'=11$, \emph{E-match} chooses \emph{Frank} as the best candidate for \emph{Alice}, and computes the exact similarity coefficients with the probability of $0.85$.

\textbf{Remark.} The range of \emph{Bob} and \emph{David} in the experiment above is different in \emph{P-match} and \emph{P-match$^+$}. This is because we consider the intersection of every two users' attributes as the sample in \emph{P-match} and the union in \emph{P-match$^+$}. That means, Tanimoto similarity ignores the number of the common attributes and it only computes the similarity of the priorities on the same attributes. Meanwhile, Ochiai similarity computes the number of the common attributes and the priorities simultaneously in the union attributes set. In the experiment above, \emph{Bob} and \emph{Alice} have 2 common attributes while \emph{David} and \emph{Alice} have 3 common attributes, but the differences on common attributes are the same. So \emph{Bob} is before \emph{David} in \emph{P-match} and after \emph{David} in \emph{P-match$^+$}.

\subsection{Complexity Analysis}
To discuss the complexity of our schemes, we analyze the online/offline computation overhead and the communication cost from both the \emph{initiator} and \emph{responder} sides. The computation cost is measured by counting the keyed hash functions and exponentiation operations, since these operations are always resource-consuming in mobile devices. $h$ represents a keyed hash function, such as SHA-256 or SHA-512, while \emph{$mul_1$}, \emph{$exp_1$}and \emph{$exp_2$} means 1024-bit multiplication, 1024-bit and 2048-bit exponentiation operations, respectively. The communication cost is evaluated by computing the transmitted and received bits. We compare our work with algorithms in \cite{Cristofaro10} and \cite{L1distance10}, since the former algorithm considers the malicious behavior in private matching as our work, and \cite{L1distance10} tries to offload the computation overhead in existing secure two-party computation, which is well used in secure private matching problem. We assume that there are several mobile users in vicinity, and each user holds $m$ attributes, where every attribute has a priority value from $[1, \kappa]$. Table \ref{tab_compare} shows the theoretic analysis in details. Our schemes have lower computation cost, especially the online parts.
\begin{table*}[!t]
\renewcommand{\arraystretch}{1.3}
\caption{Comparison of Matching Algorithms}
\label{tab_compare}
\centering
\begin{tabular}{c|c|c|c|c}
\hline
Protocols & Party & Offline Comp. & Online Comp. & Comm.trans (in bits)  \\ \hline \hline
\cite{Cristofaro10} & Initiator & $(2m+2m^2)$\emph{$exp_1$}, $(2m)$\emph{h} & $(m+m^2)$\emph{$exp_1$}, $(m)$\emph{h} & $3m\cdot1024$ \\
                    & Responder & $(m+m^2)$\emph{$exp_1$}, $(2m)$\emph{h} & $(2m)$\emph{$exp_1$} & $4m\cdot1024$ \\
\cite{L1distance10} & Initiator & $(2rm)$\emph{$exp_1$}, $(rm)$\emph{$exp_2$} & $(rm)$\emph{$exp_1$}, $(2rm)$\emph{$exp_2$} & $ rm\cdot2048$ \\
                    & Responder & ---- & $(2rm+1)$\emph{$exp_1$}, $(2rm+1)$\emph{$exp_2$} & $ rm\cdot2048$ \\
P-match & Initiator & $(2m+1)$\emph{$exp_1$}, $(m)$\emph{h} & $(2m)$\emph{$exp_1$} & $4m\cdot1024$ \\
        & Responder & $(2m+1)$\emph{$exp_1$}, $(m)$\emph{h} & $(3m)$\emph{$exp_1$} & $2m\cdot1024$ \\
E-match & Initiator & $(2m)$\emph{h}, 1\emph{poly$^+$} & ---- & 1024 \\
        & Responder & $(rm)$\emph{h}, $((r-1)m)$\emph{$mul_1$} & $(rm)$\emph{poly$^-$} & 32 \\
\hline
\end{tabular}
\end{table*}

\subsection{Experiment Setup}
To study the feasibility of our algorithms, we first evaluate the time taken for generating SHA-256, SHA-512, $exp_1$, $exp_2, $1024-bit and 2048-bit safe primes, respectively. We then implement our proposed algorithms on a Thinkpad laptop (the cryptography library is Crypto++) with 1.82 GHz CPU, 4 GB RAM, and Windows 7 32-bit Professional to simulate the performance, which is used for the offline computation. We also implement our schemes on two SAMSUNG Nexus S smartphones with 1 GHz Cortex-A8 processor, 512MB RAM, Android v2.3.6, and Bluetooth v2.1. Each result in our experiments is an average of 1000 runs.

\subsection{Experiment Results}
\label{subsec_evaluationresults}
Table \ref{tab_onpc} and \ref{tab_onmobile} show the mean, maximum, minimum, medium and standard deviation of time consumption of SHA-256, $exp_1$, $exp_2$ and generating safe primes with 1024-bit and 2048-bit, respectively. We can clearly see the better performance provided by the laptop since the powerful computing capability. For example, when we generate a 1024-bit safe prime, averagely, it needs to consume 156.37 ms on the laptop and 582.28 ms on the Nexus S, respectively. Fortunately, it needs not to be worried in our work since we can put these work in the offline computation phase.

We then show some evaluation results on the offline/online computation cost, communication cost and the execution time, respectively. Specifically, the offline communication cost means the operations which can be pre-computed without supply from other entities. The online computation cost represents the operations that to be computed in real time. The communication cost indicates the transmitted data in bits and the execution time stands for the total time consumption to perform a private matching procedure between users, including both the online computation cost and the data transmitting time between users.

\begin{table}[!t]
\renewcommand{\arraystretch}{1.3}
\caption{Time consumption of different operations on laptop}
\label{tab_onpc}
\centering
\begin{tabular}{c|c|c|c|c|c}
\hline
Operation & Mean & Max & Min & Median & Std \\
\hline
\hline
SHA-256 ($\mu$s) & 2.13 & 2.5 & 2.1 & 2.1 & 0.048  \\
SHA-512 ($\mu$s) & 10.6 & 14 & 10.5 & 10.6 & 0.21  \\
\emph{$exp_1$} ($\mu$s) & 340.6 & 483 & 338 & 339 & 7.13  \\
\emph{$exp_2$} ($\mu$s) & 756.8 & 986 & 752 & 755 & 12.69  \\
Prime-1024 (ms) & 156.37 & 178 & 134 & 156 & 12.60  \\
Prime-2048 (ms) & 1545.27 & 1663 & 1413 & 1546 & 74.26  \\
\hline
\end{tabular}
\end{table}

\begin{table}[!t]
\renewcommand{\arraystretch}{1.3}
\caption{Time consumption of different operations on Nexus S (ms)}
\label{tab_onmobile}
\centering
\begin{tabular}{c|c|c|c|c|c}
\hline
Operation & Mean & Max & Min & Median & Std \\
\hline
\hline
SHA-256 & 18.79 & 20 & 18 & 19 & 0.83  \\
SHA-512 & 22.17 & 24 & 21 & 22 & 1.10  \\
\emph{$exp_1$} & 39.17 & 75 & 20 & 39 & 9.05  \\
\emph{$exp_2$} & 59.94 & 110 & 35 & 58 & 15.31  \\
Prime-1024 & 582.28 & 650 & 525 & 582 & 37.52  \\
Prime-2048 & 7090.46 & 7175 & 6518 & 6892 & 219.36  \\
\hline
\end{tabular}
\end{table}

\subsubsection{The impact of $m$}
\begin{figure*}[!t]
  \centerline{\subfloat[]{\includegraphics[width=1.67in]{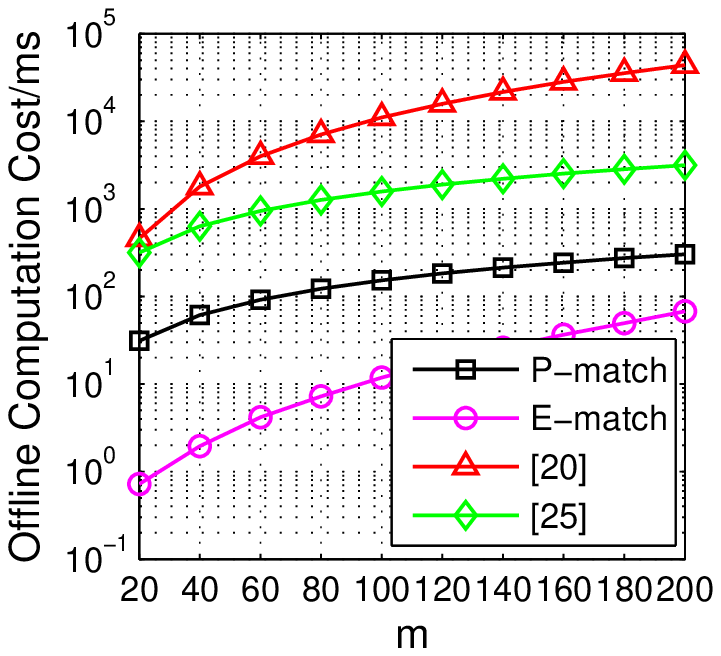}
    \label{subfig_off_m}}
    \hfil
  \subfloat[]{\includegraphics[width=1.67in]{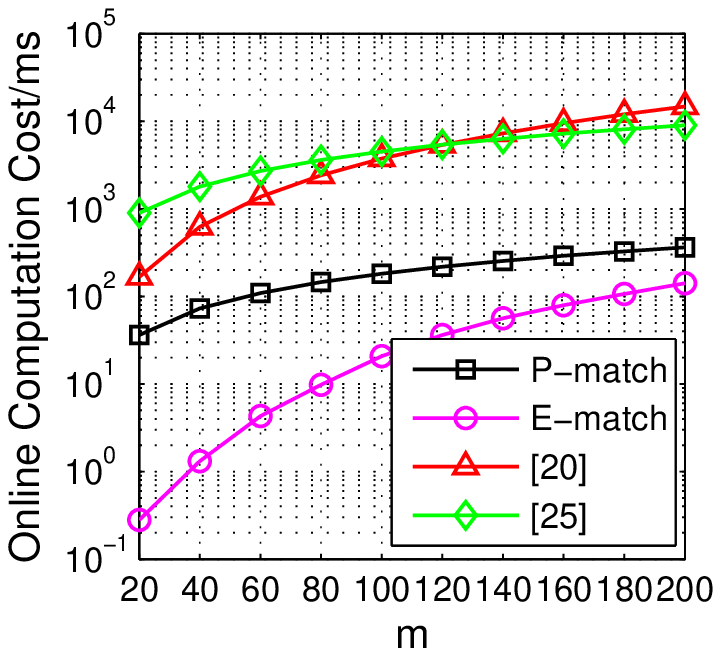}
    \label{subfig_on_m}}
    \hfil
  \subfloat[]{\includegraphics[width=1.67in]{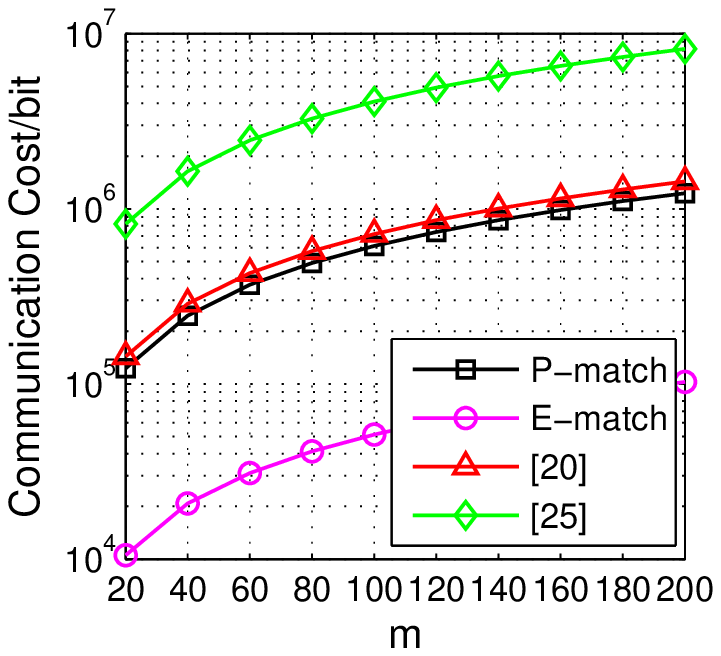}
    \label{subfig_comm_m}}
    \hfil
  \subfloat[]{\includegraphics[width=1.67in]{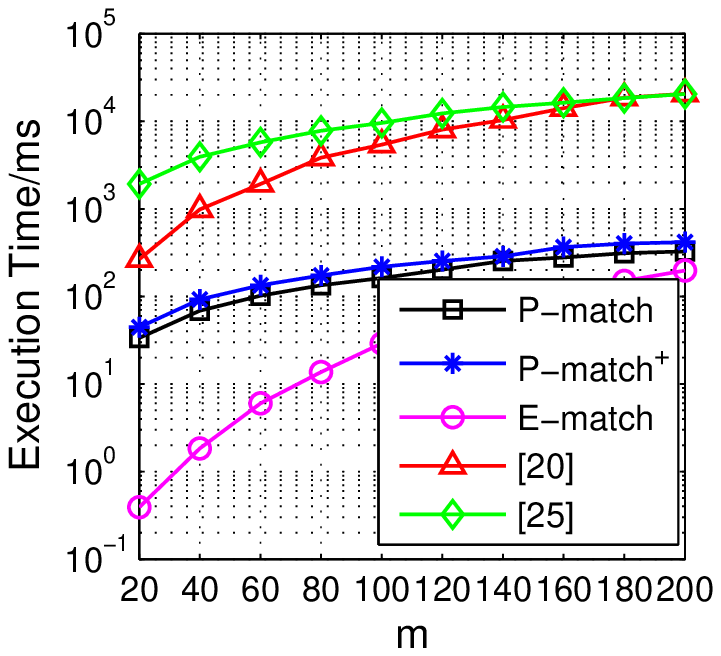}
    \label{subfig_execution_m}}}
  \caption{Impact of the number of attributes $m$ on (a): offline computation cost; (b): online computation cost; (c): communication cost; (d): the execution time, $\kappa = 10$}
  \label{fig_final_results1}
\end{figure*}

Fig. \ref{fig_final_results1} shows the evaluation results ($\kappa = 10$) of the impact of varying $m$. We first test the offline computation cost when the number of attribute $m$ is changing from 20 to 200. Fig. \ref{subfig_off_m} indicates that both of \emph{P-match} and \emph{E-match} have better performance than existing work \cite{Cristofaro10, L1distance10}. The computation cost in \cite{Cristofaro10} is high since there are too many $exp_1$s employed in their schemes. \emph{E-match} outperforms than \emph{P-match} by the reason of the utilization of \emph{poly$^-$}s instead of the $exp_1$s. The offline cost can be computed before the regular computations, so this part does not impact the execution time.

Fig. \ref{subfig_on_m} compares the online computation cost of all the protocols in the $\log10$ scale for varying $m$. It makes sense that this part is sensitive to the execution time, so we aim to offload the online computation to offline as much as possible. We can clearly see the efficiency of our protocols over others. For users in \emph{P-match}, they just need to perform $2m$ $exp_1$s on the \emph{initiator} side and $3m$ $exp_1$s on the \emph{responder} side. While in \emph{E-match}, it decreases the computation cost significantly by validating the polynomial with several potential solutions. The online cost of the protocols in \cite{Cristofaro10, L1distance10} are much higher since they utilize several $exp_1$s and $exp_2$s in their processes.

Fig. \ref{subfig_comm_m} shows the communication cost between entities. Not surprisingly, the result of each protocol increases smoothly with the increasing $m$. Our \emph{E-match} shows a better performance on the communication cost through replacing the complicated exchanging phases between entities with a single polynomial. The \emph{initiator} only needs to transmit some simple parameters of a specific polynomial to exchange both the attributes and the corresponding priorities. For example, \emph{E-match} needs to consume 50.28Kb on bandwidth when $m = 100$. This is easy for Bluetooth v2.1, since the transmission rate can achieve approximately 900Kb/s in our experiments, which means that we only need to spend 55.87 milliseconds to finish all the transmissions.

Fig. \ref{subfig_execution_m} provides the total execution time of all the algorithms. The execution time in our experiments mainly includes the online computation time and the information transmission time. Comparing with the schemes in existing work \cite{Cristofaro10, L1distance10}, our \emph{P-match}, \emph{P-match$^+$} and \emph{E-match} degrade the execution time. When we look into our protocols more specifically, to get the common attributes securely, an \emph{initiator} needs more time to complete the computation in \emph{P-match} and \emph{P-match$^+$}. However, it is obvious that all of our proposed protocols can be finished within about 600ms in all simulated sceneries. For example, when the number of attributes $m = 200$, protocols in \cite{L1distance10} and \cite{Cristofaro10} need 20.52 and 20.66 seconds to complete the matching phase for each user, while our \emph{P-match} and \emph{P-match$^+$} require 0.33 and 0.42 seconds, respectively. This value in our \emph{E-match} is 197.86 milliseconds, which is more practical for mobile users.

\subsubsection{The impact of $\kappa$}
\begin{figure*}[!t]
  \centerline{\subfloat[]{\includegraphics[width=1.67in]{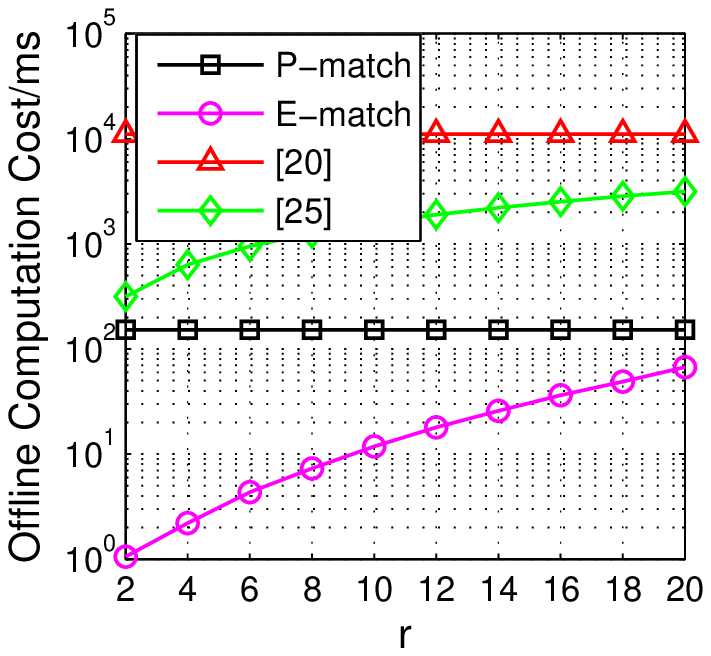}
    \label{subfig_off_r}}
    \hfil
  \subfloat[]{\includegraphics[width=1.67in]{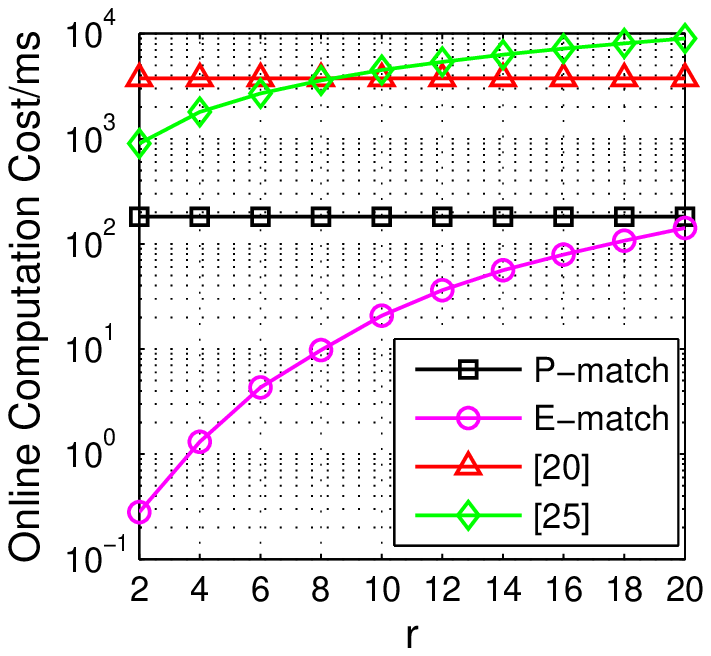}
    \label{subfig_on_r}}
    \hfil
  \subfloat[]{\includegraphics[width=1.67in]{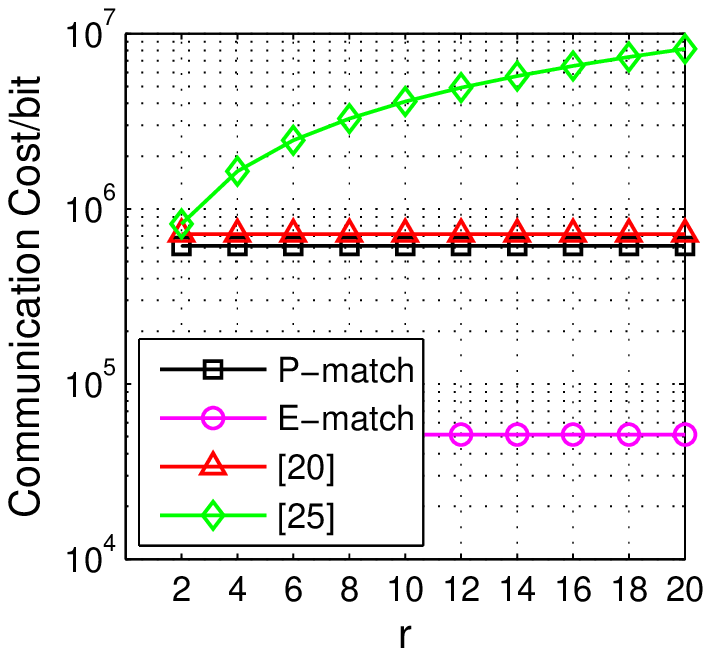}
    \label{subfig_comm_r}}
    \hfil
  \subfloat[]{\includegraphics[width=1.67in]{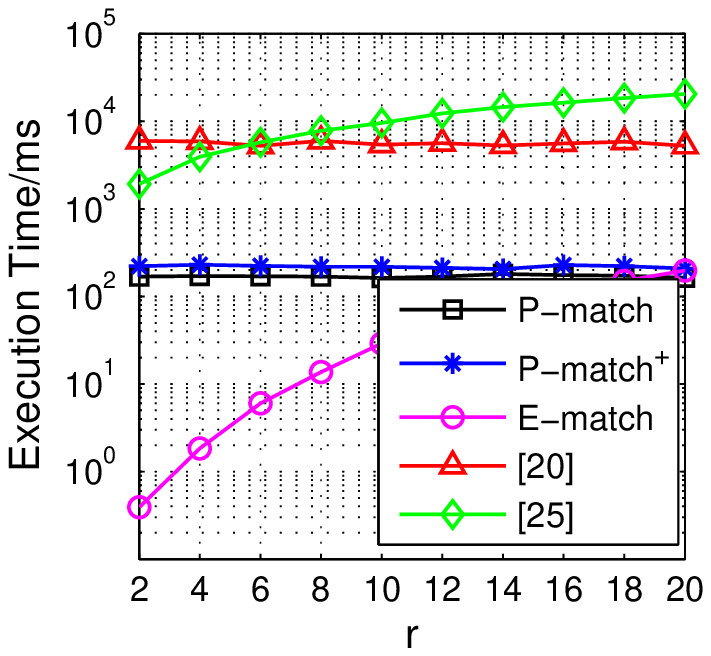}
    \label{subfig_execution_r}}}
  \caption{Impact of the upper bound of priorities $\kappa$ on (a): offline computation cost; (b): online computation cost; (c): communication cost; (d): the execution time, $m = 100$}
  \label{fig_final_results2}
\end{figure*}

In Fig. \ref{fig_final_results2}, we show some evaluation results ($m = 100$) of the impact of the varying $\kappa$. Fig. \ref{subfig_off_r} indicates the changes on the offline computation cost with the varying $\kappa$. The evaluation results show that our \emph{P-match} and \emph{E-match} outperform other schemes in all the tested $\kappa$s. \cite{Cristofaro10} and \emph{P-match} are steady in terms of various $\kappa$. Meanwhile, \cite{L1distance10} and \emph{E-match} are impacted by the changing $\kappa$. The reason is that, \cite{L1distance10} needs extra $exp_2$s to transform $\ell_1$ distance into $\ell_2$ distance by performing Johnson-Lindenstrauss embedding, and\emph{E-match} need to re-build the polynomial $f(x)$ by computing each possible priority value on the \emph{responder} side.

Fig. \ref{subfig_on_r} shows the offline computation cost of the schemes. The performance of \cite{L1distance10} is the worst one when $\kappa \geq 8$, due to a number of utilizations on the $exp_2$s while less on the $exp_1$s. \emph{E-match} outperforms than all other schemes since it did not employ the heavy operations such as $exp_1$s and $exp_2$s.


Fig. \ref{subfig_comm_r} demonstrates the general trends of the communication cost of different schemes. We can see \cite{Cristofaro10} and \emph{P-match} are stable with increasing $\kappa$. For example, when $m = 100$, the communication cost are 0.68 Mb in \cite{Cristofaro10} and 0.59 Mb in \emph{P-match}. However, in \cite{L1distance10}, it is heavily impacted by the increasing $\kappa$, the communication cost exceeds 3.91 Mb when $\kappa = 10$. This situation is changed a lot in our proposed \emph{E-match}, it is not stable with the increasing $\kappa$, nevertheless, it does not change too much, for example, the communication cost is about 50.28 Kb when $\kappa = 10$, which is very comfortable for Bluetooth v2.1.


In Fig. \ref{subfig_execution_r}, the evaluation results clearly show the advantages of our proposed schemes. For instance, when $\kappa = 10$, compared with 5.40 seconds in \cite{Cristofaro10} and 9.54 seconds in \cite{L1distance10}, our \emph{P-match} and \emph{P-match$^+$} need 161.54 and 217.81 milliseconds, respectively. And in our \emph{E-match}, the execution time is only 29.01 milliseconds to achieve the same goal.

\subsubsection{Implementation Results on SAMSUNG Nexus S Smartphones}
\label{subsec_inplementationresults}
\begin{figure*}[!t]
  \centerline{\subfloat[]{\includegraphics[width=1.67in]{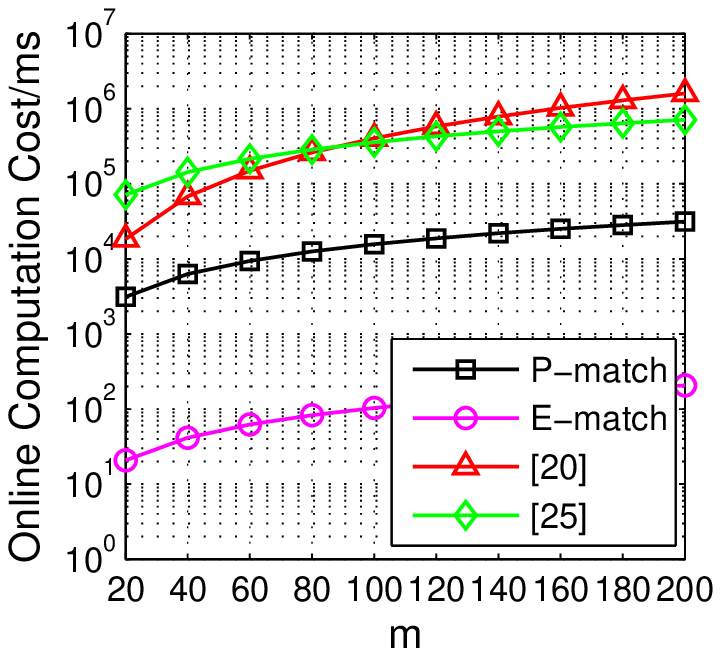}
    \label{subfig_on_m1}}
    \hfil
  \subfloat[]{\includegraphics[width=1.67in]{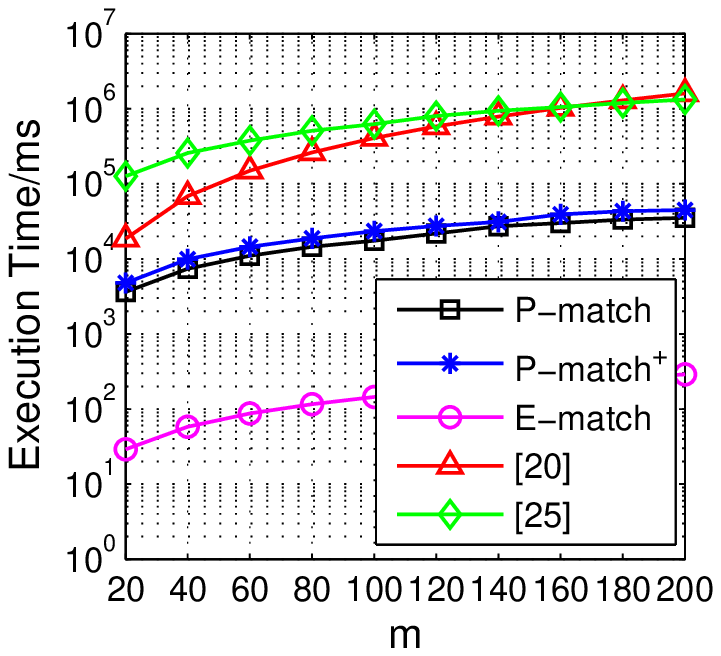}
    \label{subfig_execution_m1}}
    \hfil
  \subfloat[]{\includegraphics[width=1.67in]{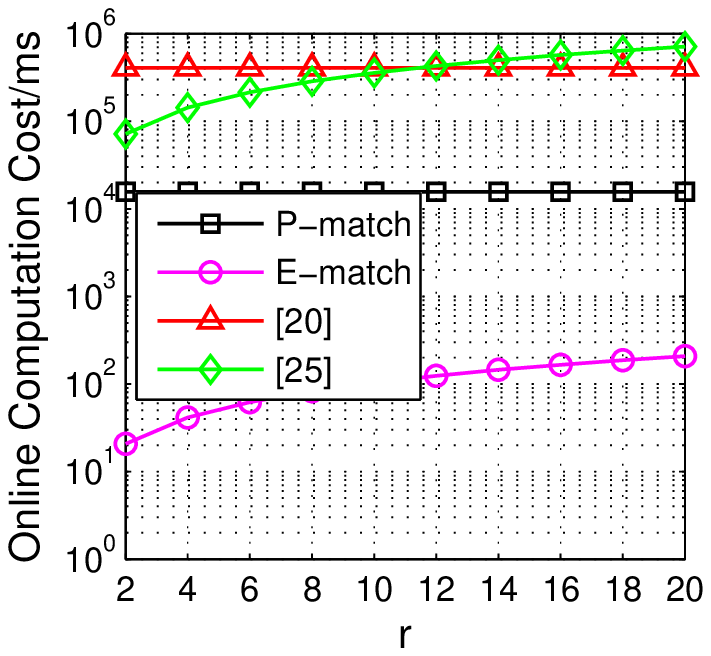}
    \label{subfig_on_r1}}
    \hfil
  \subfloat[]{\includegraphics[width=1.67in]{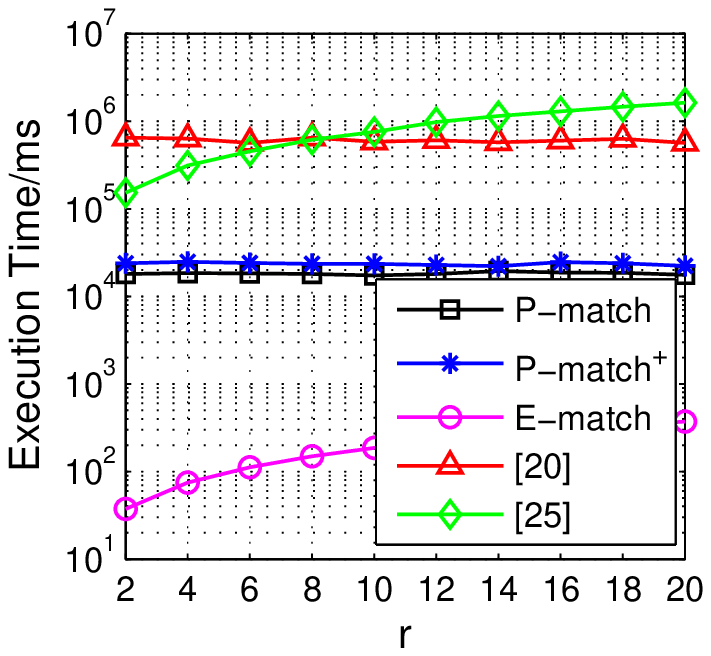}
    \label{subfig_execution_r1}}}
  \caption{Impact of number of attributes $m$ on (a): online computation cost; (b): the execution time, $\kappa = 10$, impact of the priorities $\kappa$ on (c): online computation cost; (d): the execution time, $m = 100$}
  \label{fig_final_results_nexus}
\end{figure*}

To validate the usability of our proposed protocols, we implement them on SAMSUNG Nexus S smartphones to test the performance. Fig. \ref{fig_final_results_nexus} shows some selected results on smartphones, which may be a litter different from the simulation results on laptop, the reason can be found in Table \ref{tab_onpc} and \ref{tab_onmobile}. Generally speaking, the main differences between them are the online computation cost and the execution time.

Fig. \ref{subfig_on_m1} shows the online computation cost on the Nexus S with the varying $m$. The performance of \cite{Cristofaro10} and \cite{L1distance10} are quite similar with each other, however, they need several tens of seconds or more to do the online commination, which cannot be accepted by mobile users. Our proposed \emph{P-match} and \emph{E-match} perform better than others. Specifically, \emph{E-match} is the best since it only needs to verify the polynomial by several possible results, which is quite simple for modern smartphones.

In Fig. \ref{subfig_execution_m1}, we show the changes on the execution time of different schemes. Generally, the execution time is heavily related with the online computation time, and it increases with the increasing $m$. Not surprisingly, the performance of our \emph{E-match} is much better than others.

Next, Fig. \ref{subfig_on_r1} indicates the online computation cost on the Nexus S with the changing $\kappa$ when $m$ is set to 100. Similar with the evaluation results of online computation cost on the laptop, the performance of \cite{Cristofaro10} and \cite{L1distance10} are unacceptable for mobile users since they need several minutes to finish the matching phase. Our \emph{P-match} cannot be affected by the changing $\kappa$, however, it still needs 15.67 seconds to do the matching. While in our \emph{E-match}, it is lightweight and only consumes 103.6 millisecond to complete the same phase when $\kappa = 10$.

In Fig. \ref{subfig_execution_r1}, the changes on the varying $\kappa$ bring less effect on the execution time for the smartphones in different schemes. Our \emph{E-match} outperforms than others significantly. For instance, when $\kappa = 10$, the execution time of \cite{Cristofaro10} and \cite{L1distance10} are 584.46 and 757.60 seconds, respectively. The results of our \emph{P-match} and \emph{P-match$^+$} are 17.41 and 23.47 seconds, respectively. While in our \emph{E-match}, it only needs 186.48 millisecond to process the matching with others, which is absolutely efficient.

\subsection{Energy Consumption}
We also compute and compare the energy consumption of our scheme with others. The most energy consumed operations for modern smartphones are local computation, display and network transmission \cite{Mittal12}. In our work, since we did not use heavy graphics, we pay much attention on two main factors, local computation cost and network transmission cost. We use the energy consumption model \cite{Carroll10} $E_{computing} = P_{comp} \cdot T_{comp} + 0.3167 T_{run}$ to estimate local computation cost, where $P_{comp}$ represents the CPU��s power consumption, $T_{comp}$ means the time spent for computation and $T_{run}$ indicates the total protocol run time. For a smartphone with 1 GHz CPU, we choose $P_{comp} \approx 0.38 w$ \cite{Mittal12}. The energy consumption model of the network transmission cost is based on \cite{Rahmati07}: $E_{network} = n_t \cdot E_t + n_r \cdot E_r$, where $n_t$ and $n_r$ are the transmitted and received data in bytes, and $E_t \approx 4.8 \mu J$ is transmitting energy per byte, $E_r \approx 6.7 \mu J$ is the receiving energy per byte. For simplicity, we omit the initial connection establishment energy since it is common in all schemes. Then our energy consumption model can be denoted as:
\begin{small}
\begin{eqnarray}
      \label{for_energy}
      E & = & E_{computing} + E_{network} \nonumber\\
        & = & P_{comp} T_{comp} + 0.3167 T_{run} + n_t E_t + n_r E_r.
      \end{eqnarray}
      \end{small}

\begin{figure}[!t]
  \centerline{\subfloat[$\kappa = 10$]{\includegraphics[width=1.67in]{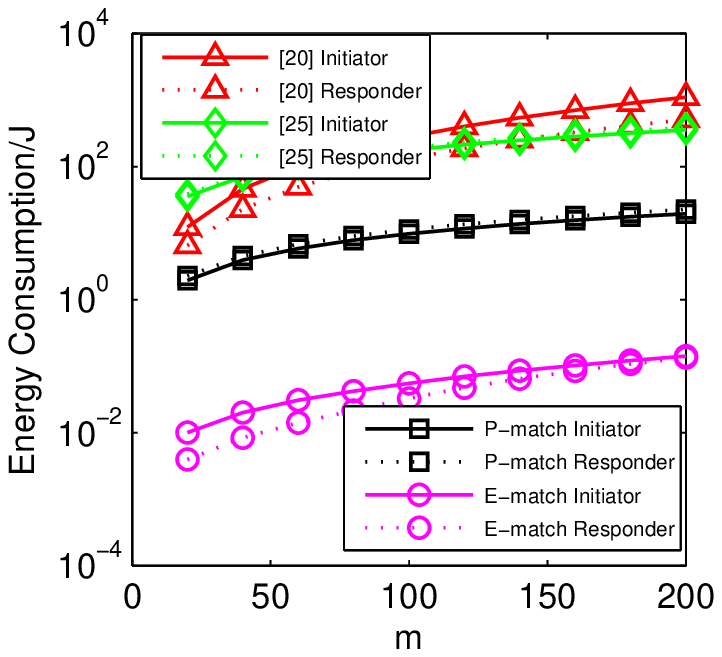}
    \label{subfig_energy_m}}
    \hfil
    \subfloat[$m = 100$]{\includegraphics[width=1.67in]{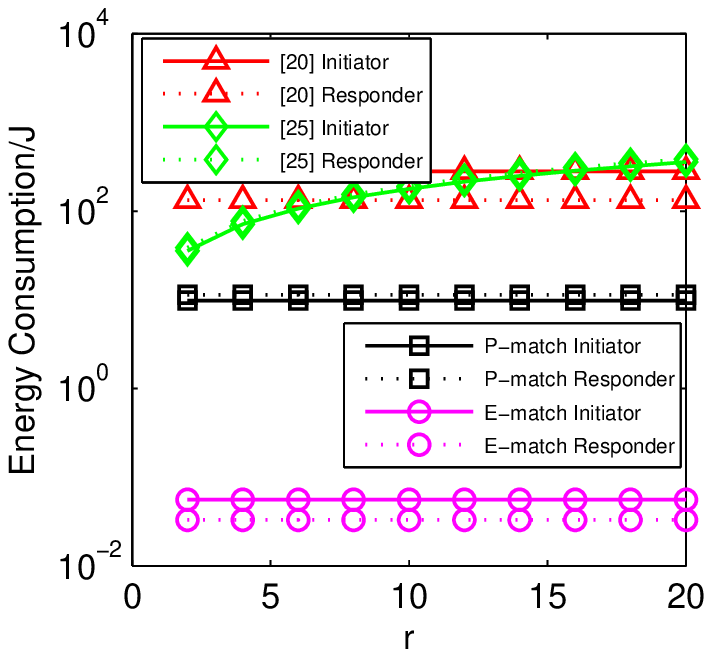}
    \label{subfig_energy_r}}}
  \caption{Energy consumption}
  \label{fig_energy}
\end{figure}

Fig. \ref{subfig_energy_m} shows the comparison of energy consumption on each side of the protocols. It is clear that our protocols consume less power than other schemes. For example, when the number of attributes $m = 100$ and $\kappa = 10$, the \emph{initiator}s in the protocols \cite{Cristofaro10, L1distance10} need to consume 280.18$J$ and 178.00$J$, respectively. While in our \emph{P-match}, it consumes lower energy of 9.81$J$ to achieve the same goal. Finally, in our efficient version \emph{E-match}, it degrades the energy consumption from both the computation and network transmission aspects, as a result, it consumes 55.60$mJ$ and 33.00$mJ$ on the \emph{initiator} and the \emph{responder} sides, respectively. We can conclude that our protocols are very practical in terms of energy consumption.

Fig. \ref{subfig_energy_r} indicates the impact of $\kappa$ on the energy consumption, the number of attributes is fixed to 100. Since the similar result with Fig. \ref{subfig_energy_m}, we only point out that the energy consumption of our \emph{E-match} is increasing slowly with the varying $\kappa$.

\section{Conclusion}
\label{sec_concl}
We propose a Priority-aware Private Matching problem to satisfy the requirements of our real social life for the first time. Comparing to existing work, the matching processes of our \emph{P-match}s consider both the number of common attributes and the corresponding priorities. To avoid possible attacks from both \emph{initiator} and \emph{responder}, we then construct a priority-aware Ochiai similarity function in our enhanced version. Finally, an efficient version called \emph{E-match} is also proposed to decrease the cost. The followed security analysis and performance evaluation show the correctness and efficiency of our algorithms. Our future work is to deploy our \emph{P-match} and \emph{E-match} into a large scale of real mobile environment to test the performance.

%

\section*{Acknowledgment}
The preliminary work is accepted by IEEE MASS 2013 \cite{Ben13}. This work was supported by National Natural Science Foundation of China under Grant 61272457.

\ifCLASSOPTIONcaptionsoff
  \newpage
\fi

\bibliographystyle{IEEEtran}
\bibliography{TDSC1}

\end{document}